\documentclass[journal]{IEEEtran}

\usepackage{graphicx}      
\usepackage{amsmath} 
\usepackage{amssymb}  
\usepackage{color,soul}
\usepackage{colortbl}
\usepackage{subfigure}
\usepackage{array}
\usepackage{booktabs}
\usepackage{url}
\usepackage{multicol}
\usepackage{epstopdf}
\usepackage{makecell}

\begin{document}
%
\title{Robust Model Predictive Control of Irrigation Systems with Active Uncertainty Learning and Data Analytics}
%
%
%

\author{Chao~Shang,~\IEEEmembership{Member,~IEEE,}
        Wei-Han~Chen,
        Abraham Duncan Stroock,
        and~Fengqi~You,~\IEEEmembership{Member,~IEEE}
\thanks{C. Shang is with Department of Automation, Tsinghua University, Beijing 100084, China (e-mail: c-shang@tsinghua.edu.cn). W. Chen, A. D. Stroock, and F. You are with Robert Frederick Smith School of Chemical and Biomolecular Engineering, Cornell University, Ithaca, New York 14853, USA (e-mail: wc593@cornell.edu; abe.stroock@cornell.edu; fengqi.you@cornell.edu). \textit{(Corresponding Author: Fengqi You. Phone: (607) 255-1162; Fax: (607) 255-9166; E-mail: fengqi.you@cornell.edu)}}
}

\maketitle

\begin{abstract}
We develop a novel data-driven robust model predictive control (DDRMPC) approach for automatic control of irrigation systems. The fundamental idea is to integrate both mechanistic models, which describe dynamics in soil moisture variations, and data-driven models, which characterize uncertainty in forecast errors of evapotranspiration and precipitation, into a holistic systems control framework. To better capture the support of uncertainty distribution, we take a new learning-based approach by constructing uncertainty sets from historical data. For evapotranspiration forecast error, the support vector clustering-based uncertainty set is adopted, which can be conveniently built from historical data. As for precipitation forecast errors, we analyze the dependence of their distribution on forecast values, and further design a tailored uncertainty set based on the properties of this type of uncertainty. In this way, the overall uncertainty distribution can be elaborately described, which finally contributes to rational and efficient control decisions. To assure the quality of data-driven uncertainty sets, a training-calibration scheme is used to provide theoretical performance guarantees. A generalized affine decision rule is adopted to obtain tractable approximations of optimal control problems, thereby ensuring the practicability of DDRMPC. Case studies using real data show that, DDRMPC can reliably maintain soil moisture above the safety level and avoid crop devastation. The proposed DDRMPC approach leads to a 40\% reduction of total water consumption compared to the fine-tuned open-loop control strategy. In comparison with the carefully tuned rule-based control and certainty equivalent model predictive control, the proposed DDRMPC approach can significantly reduce the total water consumption and improve the control performance.
\end{abstract}


\IEEEpeerreviewmaketitle

\section{Introduction}
Nowadays, the global competition for water resources is becoming more and more intense, and agriculture is under tremendous pressure from urbanization and industrialization in securing adequate water supply. It is known that about $70\%$ of water in the world is consumed for farming and irrigation purposes \cite{grafton18paradox}. What's more, up to $60\%$ of the water used for irrigation is wasted because of the inefficiency in configuring limited water resource \cite{water09united}. In this sense, when natural precipitation cannot satisfy the water needs, intelligent irrigation technologies are of vital importance in ensuring sustainability of water resource management and promoting agricultural productivity.

Formally speaking, the goal of irrigation can be deemed as maintaining the soil moisture above a prespecified level by replenishing water in a  timely and efficient manner. The recent development of sensor networks and distributed control systems provide opportunities to achieve automatic and efficient irrigations. Considerable research efforts have been made in hardware design and implementation of automatic irrigation systems \cite{kim08remote,zhao07automatic}. Another line of research concentrates on irrigation strategies. The most-used one in practice is to pre-determine the irrigation schedule based on practitioners' empirical knowledge of climatic patterns, crop demands, and soil characteristics, as commonly adopted by most commercial automatic irrigation systems. However, from the viewpoint of feedback control, this method essentially belongs to the category of \textit{open-loop control}, thereby falling short of promptly responding to deficits in soil moisture that are harmful to crop growth.

Being reactive to real-time soil moisture conditions, various closed-loop irrigation methods have been developed and implemented in the past decades. The simplest ones are referred to as ``rule-based" or ``ON-OFF" strategies, in which a certain amount of water will be supplied once soil moisture deficiency is detected. However, parameters are generally tuned by practitioners in an empirical manner, and hence model information is not systematically utilized, which erect obstacles for further improvement of control performance.

As a prevalently applied model-based control technology, model predictive control (MPC) can optimize the predicted future system behavior under explicit constraints and derives the optimal control sequence at every instant \cite{qin03survey,rawlings09model,chu2015model}. After implementing the first input, the controller moves to the next time step and solves the optimization problem repeatedly. In this way, MPC achieves real-time response to the variations of system states and inherent robustness to random disturbances. Applications of MPC in irrigation systems have emerged due to its capability of explicitly handling constraints on soil moisture levels and irrigation amounts \cite{mccarthy14simulation,delgoda16irrigation}. Meanwhile, various forecasts, including evapotranspiration and precipitation forecasts, can be easily incorporated in the MPC paradigm, which are potentially helpful for reducing water consumptions and improving efficiency of irrigation systems. However, due to sensor imperfections, unknown disturbances and limited capabilities of forecasting technologies, forecasts themselves are inevitably prone to uncertainty, which exerts significant influence on control performance and finally affects crop productivity. Unfortunately, this issue is not systematically investigated by existing MPC-based irrigation control strategies, and it leaves much room to further enhance the efficiency of automatic irrigation systems.

To develop efficient closed-loop irrigation control technologies, there are several research challenges that must be critically addressed. The first challenge is how to utilize abundant historical data to appropriately characterize the uncertainty within various forecasts, since analytic uncertainty models are typically difficult to establish in practice. The second one is how to effectively integrate mechanistic models built on domain-specific knowledge, with data-driven models built based on machine learning and data analytics, into a holistic robust MPC (RMPC) framework to derive rational control decisions. The last conundrum is how to solve the induced optimal control problems with acceptable computational costs, such that the framework can be handily implemented in practice.

To systematically address these issues, we propose in this paper a novel data-driven RMPC (DDRMPC) framework for efficient irrigation control. We use water balance models to describe the dynamics of soil moisture level, and in order to make full use of real-time forecasts, tailored data-driven uncertainty sets are developed to accurately describe the distribution of uncertain forecast errors. In this way, information of both first-principle mechanisms and historical data can be synthesized \cite{rosolia18data,ning17data}. The uncertain prediction errors are decomposed into two parts, one for evapotranspiration and the other one for precipitation. A support vector clustering (SVC)-based uncertainty set, which can be actively constructed from uncertainty data \cite{shang17data}, is adopted to describe the distribution of evapotranspiration forecast error, thereby seamlessly incorporating historical data information into RMPC. Then, by analyzing the distributional characteristics of precipitation forecast errors, we find that their distributions critically depend on the forecast values, and further propose a specific conditional uncertainty set to address such dependence in a data-driven manner. In this way, the overall uncertainty distribution can be elaborately described using historical data information, which finally contributes to rational control decisions and improvements in control performance. To ensure the quality of data-driven uncertainty sets, we further endow uncertainty sets with appropriate probabilistic guarantees by using a training-calibration procedure. The generalized affine decision rule (GADF) is then employed to provide tractable approximations to the induced optimization problems, which can be solved conveniently using off-the-shelf solvers and hence secure the ease of practical implementation of DDRMPC. Finally, we carry out comprehensive simulations based on real weather condition data to demonstrate the effectiveness of the proposed DDRMPC framework for irrigation control.


\textit{Notations and Definitions:} $\mathbb{N}_0$ is the set of non-negative natural numbers, and the set of consecutive non-negative integers $\{j,\cdots,k\}$ is denoted by $\mathbb{N}_{j:k}$. The $p$-norm of a matrix is denoted by $\|\cdot\|_p$, while $\|\cdot\|$ represents the Euclidean norm of a vector by convention. The Kronecker product operator is denoted by $\otimes$. Denote by $\bold{I}$ and $\bold{0}$ the identity matrix and the zero vector with dimensions deemed obviously by context. The Minkowski sum of two sets $\mathcal{U}$ and $\mathcal{V}$ is defined as $\mathcal{U} + \mathcal{V} = \{ \bold{u} + \bold{v} | \bold{u} \in \mathcal{U}, \bold{v} \in \mathcal{V} \}$.

\section{Preliminaries}
\subsection{Dynamic Model Based on Water Balance}

\begin{figure}[t]
\centering
\includegraphics[width = 0.4\textwidth]{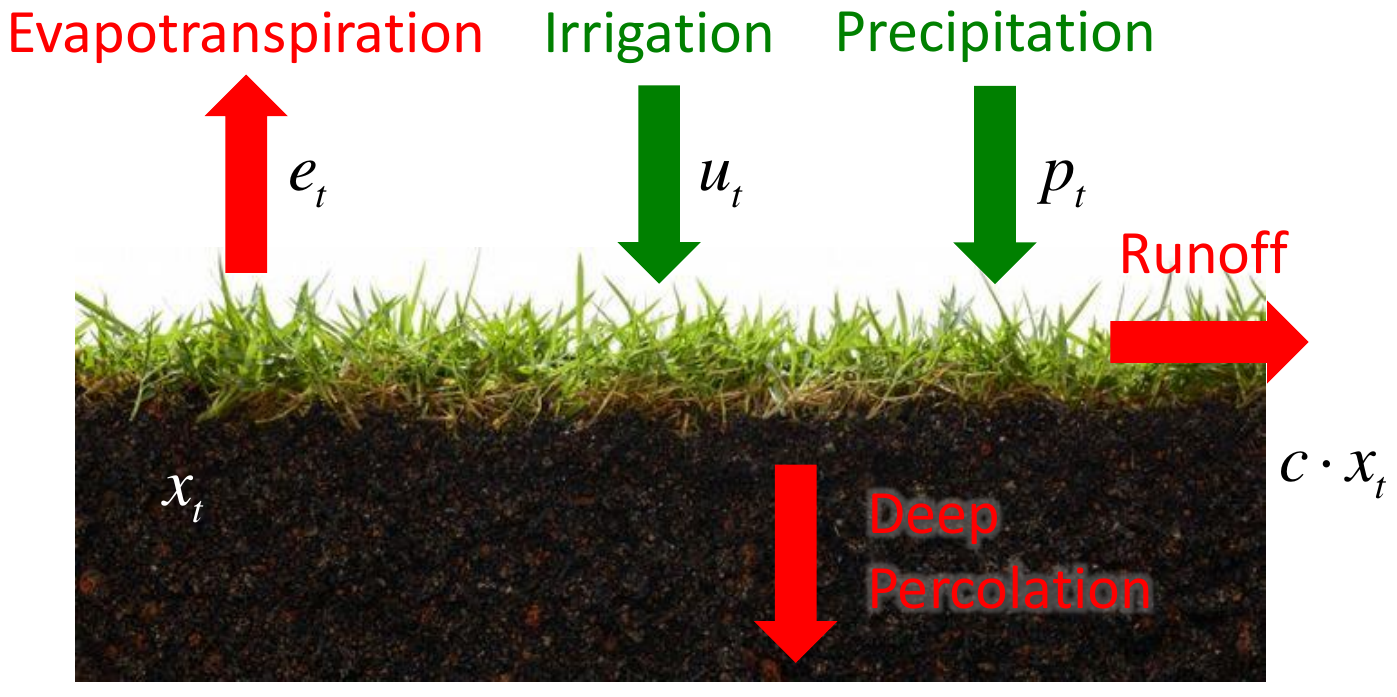}
\caption{Water balance in the root zone soil with variables defined in (\ref{eq:state_space}).}
\label{fig:waterbalance}
\end{figure}

We use a simplified first-order auto-regressive model to describe the water balance in the root zone soil \cite{delgoda16irrigation}. As shown in Fig. \ref{fig:waterbalance}, variations of water storage in the soil are caused by both inflows (irrigation and precipitation) and outflows (evapotranspiration, runoff and deep percolation). This leads to the following mathematical expression:
\begin{equation}
x_{t+1} = (1 - c) \cdot x_{t} + u_t - e_t + p_t,
\label{eq:state_space}
\end{equation}
where $x_t$ denotes the amount of water in soil, $e_t$ and $p_t$ are the cumulative evapotranspiration and precipitation in time period $t$, and $u_t$ is the irrigation amount. It is assumed in this work that runoff and water percolation are proportional to soil moisture level \cite{ooi08systems}, modelled by a decay factor $0 < c < 1$ in (\ref{eq:state_space}), where the value of $c$ can be determined by means of system identification techniques \cite{delgoda16irrigation}. All variables are normally expressed in millimeters (mm). If quantitative predictions for evapotranspiration and precipitation are available, the water balance equation (\ref{eq:state_space}) can be further rewritten as:
\begin{equation}
x_{t+1} = (1 - c) \cdot x_{t} + u_t - \hat{e}_t - \eta_t + \hat{p}_t + \xi_t,
\end{equation}
where $\hat{e}_t$ and $\hat{p}_t$ denote predictions of evapotranspiration and precipitation in time period $t$, respectively, and $\eta_t$ and $\xi_t$ are the associated prediction errors. Therefore, it can be formally expressed as the following discrete-time LTI system:
\begin{equation}
x_{t+1} = Ax_t + B_u u_t + B_v v_t + B_w w_t,
\label{eq3}
\end{equation}
where $A = 1 - c$ and $B_u = B_v = B_w = 1$. The external input $v_t = \hat{p}_t - \hat{e}_t$ is introduced to account for evapotranspiration and precipitation forecasts that are known \textit{a priori}, while $w_t = \xi_t - \eta_t$ denotes the additive disturbance arising from forecast errors, whose distribution is not necessarily known. Given the length $H$ of prediction horizon, a state-space expression of the water balance model (\ref{eq3}) is given by:
\begin{equation}
\bold{x} = \bold{A}x_0 + \bold{B}_u \bold{u} + \bold{B}_v \bold{v} + \bold{B}_w \bold{w},
\label{eq:linearsystems}
\end{equation}
where $\bold{A}$, $\bold{B}_u$, $\bold{B}_v$ and $\bold{B}_w$ are appropriately structured dynamics matrices. The model (\ref{eq:linearsystems}) enjoys considerable generality with wide applications in building control \cite{oldewurtel14stochastic} and smart-grid operations \cite{zong12application}, where the uncertainty in predictions of available energy intensity is considered to affect the dynamic system. Furthermore, it is assumed that the pairs $(\bold{A},\bold{B}_u)$ are stabilizable. State, input and disturbance sequences are defined as:
\begin{gather}
\bold{x} = \left [ x_0^{\rm T}\ \cdots\ x_H^{\rm T} \right ]^{\rm T},\ \bold{u} = \left [ u_0^{\rm T}\ \cdots\ u_H^{\rm T} \right ]^{\rm T}, \\
\bold{v} = \hat{\bold{p}} - \hat{\bold{e}},\ \hat{\bold{p}} = \left [ \hat{p}_0^{\rm T}\ \cdots\ \hat{p}_H^{\rm T} \right ]^{\rm T}, \ \hat{\bold{e}} = \left [ \hat{e}_0^{\rm T}\ \cdots\ \hat{e}_H^{\rm T} \right ]^{\rm T}, \\
\bold{w} = \boldsymbol{\xi} - \boldsymbol{\eta},\ \boldsymbol{\xi} = \left [ \xi_0^{\rm T}\ \cdots\ \xi_H^{\rm T} \right ]^{\rm T}, \ \boldsymbol{\eta} = \left [ \eta_0^{\rm T}\ \cdots\ \eta_H^{\rm T} \right ]^{\rm T}.
\end{gather}
Different from generic settings of RMPC, we do not enforce the stochastic disturbance $w_t$ in (\ref{eq3}) to be i.i.d., because of its inconsistency with the physical truth in irrigation systems. Rather, our focus is directly placed on the entire disturbance sequence $\bold{w}$, whose possible realizations are described by a bounded uncertainty set $\mathcal{W} \subseteq \mathbb{R}^H$, which includes the origin in its interior.

During the control horizon, we enforce the soil moisture to be above a certain level, that is, $x_t \ge x_{\min}$, which can be interpreted as a series of polytopic state constraints:
\begin{equation}
F_x x_t \le f_x,\ t \in \mathbb{N}_{1:H},
\end{equation}
where $F_x = -1$, $f_x = -x_{\min}$, and $x_{\min}$ is the minimum soil moisture level. Such state constraints are introduced with the aim to maintain an appropriate amount of water in the soil. This control is important because moisture deficits in soil influence crop growth, yield, quality and susceptibility to disease \cite{cohen07hydraulic,fermandez97a,fermandez97b}. Note that upper-bounds of soil moisture levels can be also included in this scheme to avoid excessive watering \cite{mao18soil}. In addition, polytopic input constraints are also considered to describe practical limitations of water supply:
\begin{equation}
F_u u_t \le f_u,\ t \in \mathbb{N}_{1:H},
\end{equation}
where $F_{u} = [1\ -1]^{\rm T}$, $f_u = [u_{\max}\ 0]^{\rm T}$ and $u_{\max}$ is the maximum irrigation amount in each time interval. By stacking all constraints in the control horizon together, we can arrive at the following succinct expressions:
\begin{equation}
\bold{F}_x \bold{x} \le \bold{f}_x,\ \bold{F}_u \bold{u} \le \bold{f}_u,
\label{eq:constraints}
\end{equation}
where $\bold{F}_x = \bold{I} \otimes F_x$ and $\bold{F}_u = \bold{I} \otimes F_u$. It is worth mentioning that more general constraints related to economic and environmental performances can be handled within the RMPC framework. For instance, constraints can be imposed directly on the quality of some high-valued crops, which is a physiological function of soil moisture levels.

\subsection{RMPC}
RMPC is typically adopted to ensure constraint satisfaction under bounded disturbances, with extensive applications in various fields \cite{saltik18outlook}. In the receding horizon implementation of RMPC of the linear system (\ref{eq:linearsystems}), system inputs $\bold{u}$ are essentially determined in response to random disturbances $\bold{w}$, thereby being uncertain as well. As a consequence, we wish to design a causal disturbance feedback control policy $\boldsymbol{\pi}(\bold{w}) := [\pi_0(\cdot), \cdots, \pi_{H-1}(\cdot)]$, where the causality requires $\pi_t(\cdot)$ to depend only on the disturbance sequence $(w_0,\cdots,w_{t-1})$ that is realized before time period $t$. The optimal policy ensures constraint satisfaction for all possible realizations of uncertainty $\bold{w} \in \mathcal{W}$:
\begin{equation}
\bold{F}_x \bold{x} \le \bold{f}_x,\ \bold{F}_u \boldsymbol{\pi}(\bold{w}) \le \bold{f}_u,\ \forall \bold{w} \in \mathcal{W},
\label{eq:robustconstraints}
\end{equation}
where the bounded set $\mathcal{W}$ is typically called the uncertainty set. Given the initial state $x_0 = x$ and a properly defined cost function $J_H(x, \boldsymbol{\pi})$, the optimization problem of RMPC can be formally expressed as:
\begin{equation}
\begin{split}
\min_{\boldsymbol{\pi}} &\ J_H(x, \boldsymbol{\pi}) \\
{\rm s.t.} &\ \bold{x} = \bold{A}x + \bold{B}_u \boldsymbol{\pi}(\bold{w}) + \bold{B}_v \bold{v} + \bold{B}_w \bold{w} \\
&\ \bold{F}_x \bold{x} \le \bold{f}_x,\ \forall \bold{w} \in \mathcal{W} \\
&\ \bold{F}_u \boldsymbol{\pi}(\bold{w}) \le \bold{f}_u,\ \forall \bold{w} \in \mathcal{W} \\
\end{split}
\label{eq:RMPC}
\end{equation}
which is solved in a receding horizon fashion. There are a number of options for the cost function $J_H(x, \boldsymbol{\pi})$ \cite{saltik18outlook}, including the nominal cost:
\begin{equation}
J_H(x, \boldsymbol{\pi}) = \sum_{t = 0}^{H-1} l(\bar{x}_t, \bar{u}_t) + l_f(\bar{x}_H),
\label{eq:nominal_cost}
\end{equation}
the worst-case cost:
\begin{equation}
J_H(x, \boldsymbol{\pi}) = \max_{\bold{w} \in \mathcal{W}} \sum_{t = 0}^{H-1} l(x_t, \pi_t(\bold{w})) + l_f(x_H),
\end{equation}
and the expected cost:
\begin{equation}
J_H(x, \boldsymbol{\pi}) = \mathbb{E} \left \{ \sum_{t = 0}^{H-1} l(x_t, \pi_t(\bold{w})) + l_f(x_H) \right \},
\label{eq:expectedcost}
\end{equation}
where $l(x,u)$ is the cost-to-go loss function of system states and inputs, and $l_f(x)$ is the terminal loss function. In the nominal cost (\ref{eq:nominal_cost}), $\{ \bar{x}_t \}$ and $\{ \bar{u}_t \}$ denote nominal predicted trajectories of system states and inputs in the absence of disturbance, that is, $\bold{w} = \bold{0}$. By deliberately designing loss functions in RMPC, one can pursue goals of minimizing water usage by penalizing large values of $\{ u_t \}$, and/or maintaining soil moisture around a prespecified level according to other economic criteria. Meanwhile, the control horizon could be several days or several weeks depending on the forecast model \cite{chen14post,kirtman14north,schepen18bayesian}. When the forecast horizon does not match the decision horizon, the terminal loss function provides more flexibility to drive the system towards a preferable state beyond the horizon. For instance, more weights can be put on $l_f(\cdot)$ by the user to account for possibility of sufficient water supply in future.

\section{Uncertainty Set Formulations}
The uncertainty set $\mathcal{W}$ of stochastic disturbance sequence $\bold{w}$ plays an important role in RMPC, since it exerts direct influence on the conservatism of control decisions. If the size of $\mathcal{W}$ is excessively large, the attained control decision has to protect against unrealistic extreme uncertainty, which leads to poor control performance in the nominal case. Conversely, if the size of $\mathcal{W}$ is small, insufficient protection is provided and there could be constraint violations. Therefore, in order to obtain satisfactory performance of RMPC, the uncertainty set must be carefully devised to capture the distributional characteristics of disturbance. An implicit assumption is that the distribution of short-term prediction errors remains approximately unaltered, although there exists variability in the distribution of weather conditions in different years. This is also a standing assumption and an observation in robust operations of energy systems, such as reservoir management \cite{Schwanenberg15short}, building climate control \cite{oldewurtel14stochastic} and smart grid operations \cite{zhao13unified,ning19data}, where invariant distributions of short-term prediction errors of energy intensities are described directly with historical scenarios collected in recent years. Note that in the state-space model (\ref{eq:linearsystems}), the uncertain disturbance sequence $\bold{w} = \boldsymbol{\xi} - \boldsymbol{\eta}$ is the difference between the forecast error of precipitation $\boldsymbol{\xi}$, and that of evapotranspiration $\boldsymbol{\eta}$. In this section, therefore, we seek to design individual uncertainty sets for $\boldsymbol{\eta}$ and $\boldsymbol{\xi}$ based on machine learning and in-depth data analytics, which finally constitute the expression of $\mathcal{W}$.

\subsection{Uncertainty Description of Evapotranspiration Forecast Error}
To build an uncertainty set for the sequence of evapotranspiration forecast error $\boldsymbol{\eta}$, a prediction model must be available to derive $\hat{\bold{e}}$. A variety of models have been proposed in the literature, which typically utilize sunlight intensity, wind speed, relative humidity and temperature as input variables \cite{vij10assessing}. Assume that we have collected $N$ sample pairs of evapotranspiration predictions $\{ \hat{\bold{e}}^{(i)} \}_{i=1}^N$ derived based on a certain forecast model, and the associated measurements $\{ \bold{e}^{(i)} \}_{i=1}^N$. Then, a set of $N$ data samples $\{ \boldsymbol{\eta}^{(i)} = \bold{e}^{(i)} - \hat{\bold{e}}^{(i)} \}_{i=1}^N$ representing uncertainty in evapotranspiration forecast errors can be obtained. The goal is then to actively estimate the support $\mathcal{D}_{\boldsymbol{\eta}}$ of $\boldsymbol{\eta}$ from training samples $\{ \boldsymbol{\eta}^{(i)} \}_{i=1}^N$, which serves as the \textit{data-driven uncertainty set} in RMPC. This is achieved by employing SVC based on weighted generalized intersection kernel (WGIK), which was proposed in \cite{shang17data} as a data-driven uncertainty set specifically tailored to robust optimization, as shown below:
\begin{equation}
\boldsymbol{\eta} \in \mathcal{D}_{\boldsymbol{\eta}} = \left \{ \boldsymbol{\eta} \left | \sum_{i \in {\rm SV}} \alpha_i \| \bold{Q}(\boldsymbol{\eta} - \boldsymbol{\eta}^{(i)}) \|_1 \le \theta \right . \right \},
\label{eq:SVCset}
\end{equation}
where $\bold{Q}$ is a weighting matrix. Model parameters $\{ \alpha_i \}$ and uncertainty set parameters $\theta$ can be determined by solving a quadratic program (QP). The attractiveness of the data-driven uncertainty set $\mathcal{D}_{\boldsymbol{\eta}}$ lie in that, the asymmetry and correlation in the uncertainty distribution can be well described by using SVC, and outliers in training samples $\{ \boldsymbol{\eta}^{(i)} \}_{i=1}^N$ can be seamlessly tackled; besides, (\ref{eq:SVCset}) is essentially a polytope, which turns out to underpin the tractability of the induced robust optimization problem. More details of the SVC-based uncertainty set are included in the supplementary document.

\subsection{Uncertainty Description of Precipitation Forecast Error}
In general, future precipitation predictions are directly available from weather forecasts. By collecting forecast values $\{ \hat{\bold{p}}^{(i)} \}_{i=1}^N$ and the corresponding measurements $\{ \bold{p}^{(i)} \}_{i=1}^N$, realizations of precipitation forecast errors $\{ \boldsymbol{\xi}^{(i)} = \bold{p}^{(i)} - \hat{\bold{p}}^{(i)} \}_{i=1}^N$ can be readily obtained. One could also adopt ensemble climate forecasts to enhance the accuracy of $\hat{\bold{p}}^{(i)}$ by synthesizing information from multiple models into a single forecast trajectory, which have become increasingly available to water management \cite{kirtman14north,schepen18bayesian,tian12forecasting}. Due to limited data availability, precipitation forecasts from a single model are employed in this work.

\begin{figure*}[htb]
  \centering
  \subfigure[6-hour ahead ($t = 1$)]{
    \includegraphics[width=0.25\textwidth]{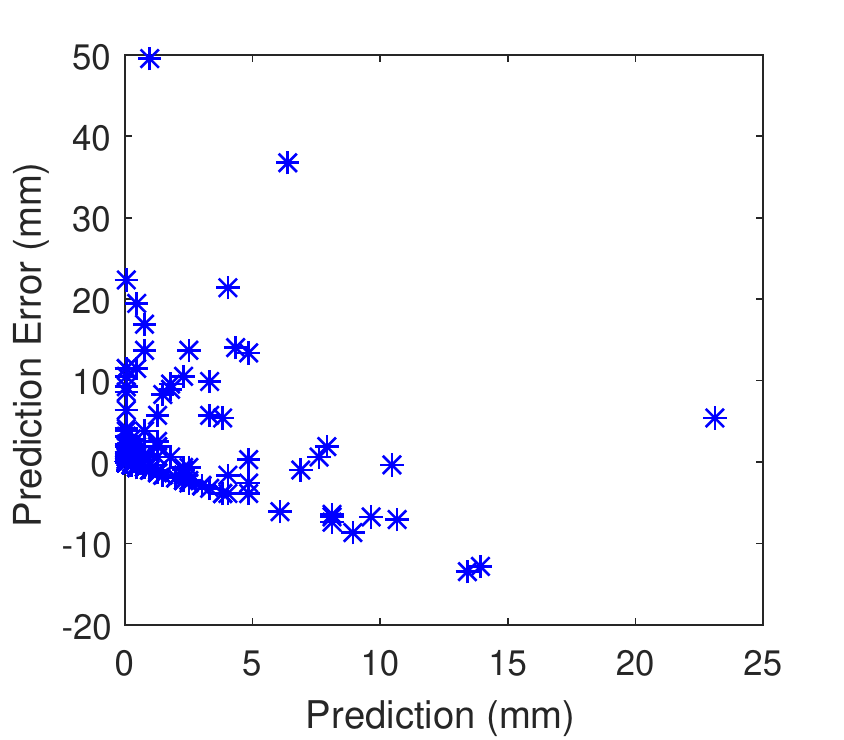}}
  \subfigure[12-hour ahead ($t = 2$)]{
    \includegraphics[width=0.25\textwidth]{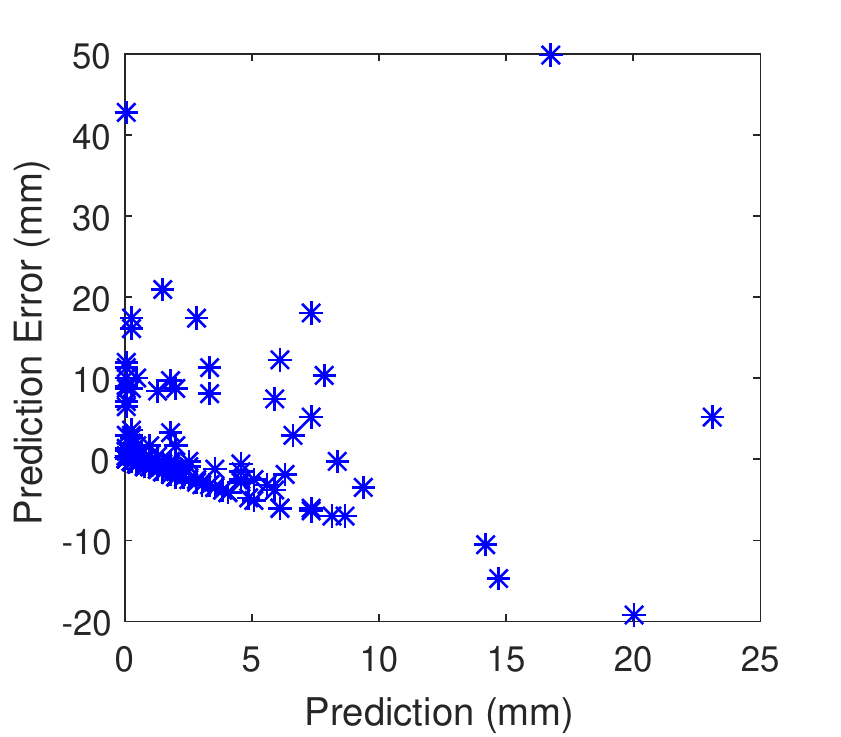}}
  \subfigure[18-hour ahead ($t = 3$)]{
    \includegraphics[width=0.25\textwidth]{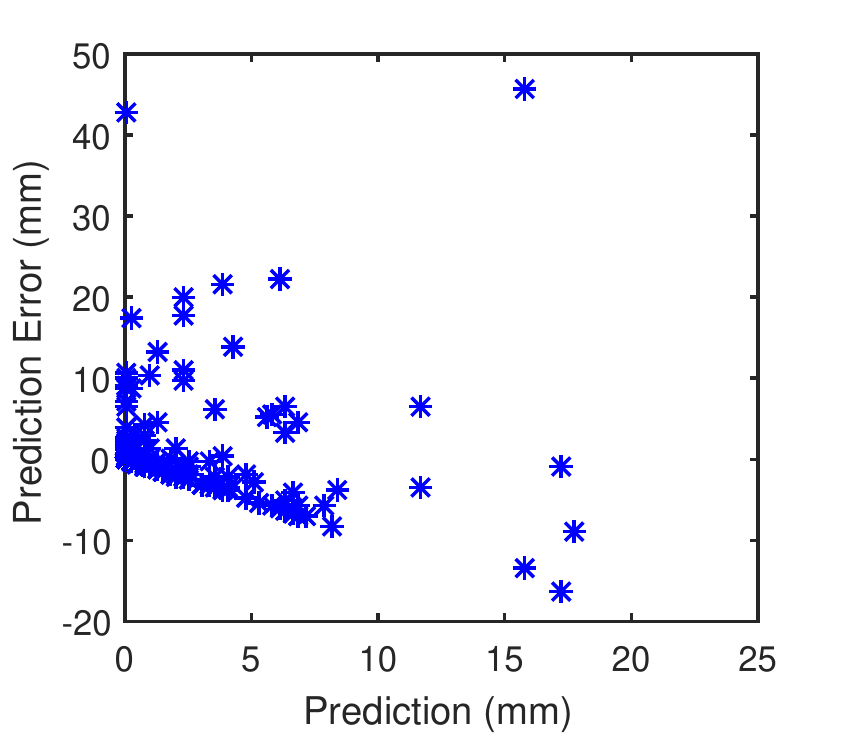}}
  \subfigure[24-hour ahead ($t = 4$)]{
    \includegraphics[width=0.25\textwidth]{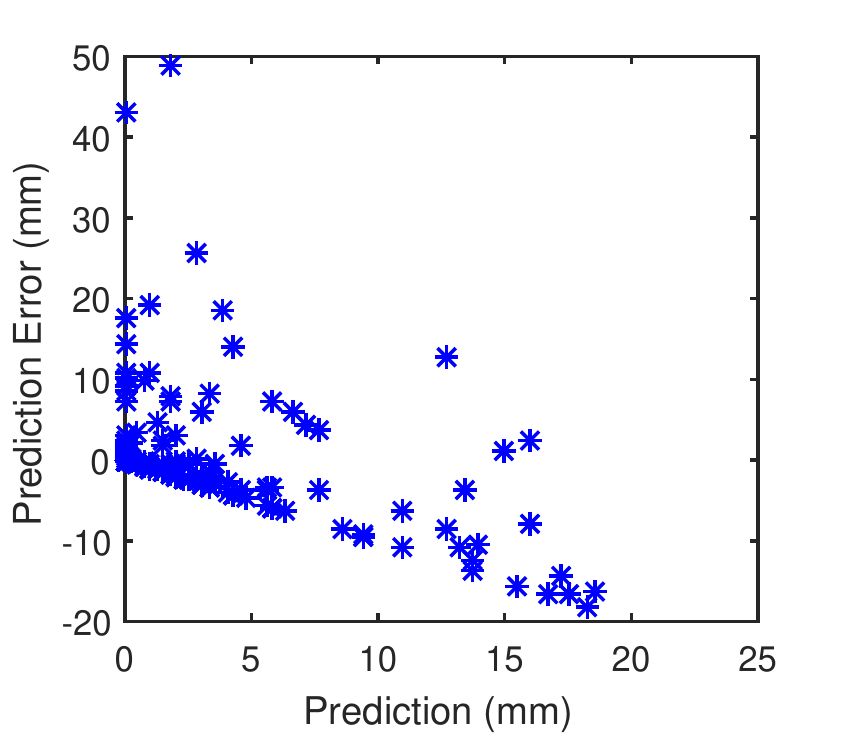}}
  \subfigure[30-hour ahead ($t = 5$)]{
    \includegraphics[width=0.25\textwidth]{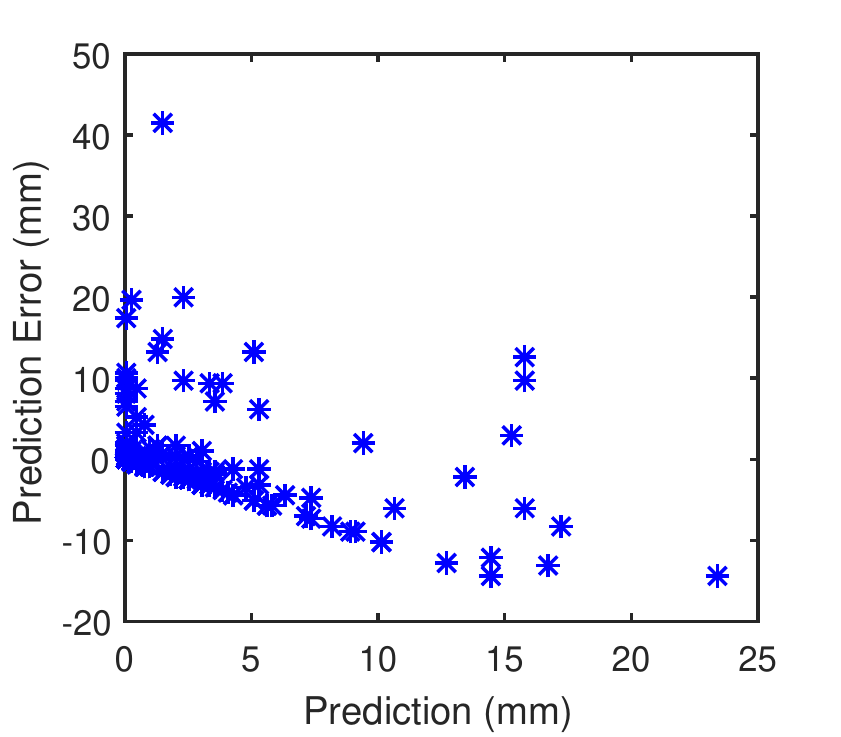}}
  \subfigure[36-hour ahead ($t = 6$)]{
    \includegraphics[width=0.25\textwidth]{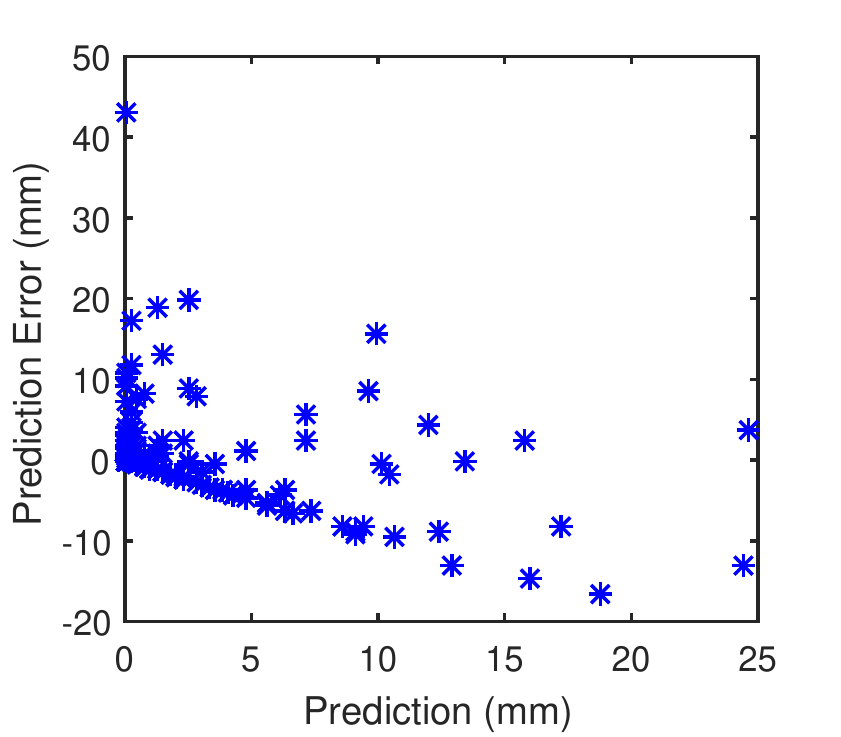}}
  \caption{Prediction errors versus prediction values of cumulative precipitation at Des Moines, Iowa from May 2016 to Oct. 2016 \cite{iowa}.}
  \label{fig:Prediction_vs_Error} 
\end{figure*}

Different from the case of evapotranspiration, however, one cannot directly apply the SVC-based uncertainty set to estimate the support of $\boldsymbol{\xi}$, primarily because precipitation is essentially a discrete event, and its intensity is always non-negative. In Fig. \ref{fig:Prediction_vs_Error}, the empirical distribution of $\boldsymbol{\xi}$ and $\hat{\bold{p}}$ is depicted, which is based on real precipitation forecast and forecast error data collected at Des Moines, Iowa in 2016 \cite{iowa}. The forecast data typically have the following properties.

On one hand, the distribution of forecast error $\xi_t$ is apparently \textit{asymmetric} around its nominal value, which is zero. A generic approach to handle asymmetry in the robust optimization is to decompose the uncertainty into forward deviation $\boldsymbol{\xi}^+$ and backward deviation $\boldsymbol{\xi}^-$ \cite{chen07robust,ning18data}:
\begin{equation}
\begin{split}
\boldsymbol{\xi} = \bold{C}\boldsymbol{\xi}^+ - \bold{D}\boldsymbol{\xi}^-, \bar{\boldsymbol{\xi}} = \boldsymbol{\xi}^+ - \boldsymbol{\xi}^- \in \Xi,\ \boldsymbol{\xi}^+,\boldsymbol{\xi}^- \ge \bold{0},
\end{split}
\label{eq:asymset}
\end{equation}
where $\bold{C} = \text{diag}\{c_1,\cdots,c_H\}$ and $\bold{D} = \text{diag}\{d_1,\cdots,d_H\}$ are diagonal matrices including scaling parameters. The discrepancy between $\bold{C}$ and $\bold{D}$ is the main cause for the asymmetry in $\boldsymbol{\xi}$. Forward and backward deviations $\boldsymbol{\xi}^+$ and $\boldsymbol{\xi}^-$ can be interpreted as being normalized to have the same scale. Therefore, the difference $\bar{\boldsymbol{\xi}} = \boldsymbol{\xi}^+ - \boldsymbol{\xi}^-$ tends to have a homogeneous distribution, which can be well characterized by some generic uncertainty set $\Xi$, e.g. the norm-based uncertainty set \cite{chen07robust}. However, (\ref{eq:asymset}) cannot be directly employed to describe the distribution of precipitation forecast error $\boldsymbol{ \xi }$, which varies significantly with different values of $\hat{p}_t$, as indicated by Fig. \ref{fig:Prediction_vs_Error}. Specifically, when the value of forecast error is negative, i.e. $\xi_t < 0$, the realistic precipitation amount will be smaller than the forecasted value, that is, $p_t < \hat{p}_t$. In this case, the real precipitation amount must be nonnegative, i.e. $p_t \ge 0$, and the worst-case is that there is no rainfall at all, i.e. $p_t = 0$. This indicates that the lower-bound of $\xi_t$ depends on $\hat{p}_t$:
\begin{equation}
\xi_t = p_t -\hat{p}_t \ge -\hat{p}_t.
\label{eq:lowerdependence}
\end{equation}

Next, we state that the dependence holds for the upper-bound as well. Typically, there is a maximum precipitation amount $p_{\max}$ that the realistic precipitation amount $p_t$ cannot exceed in a period, that is, $p_t \le p_{\max}$, which implies:
\begin{equation}
\xi_t = p_t -\hat{p}_t \le p^{\max} - \hat{p}_t.
\label{eq:upperdependence}
\end{equation}
Eq. (\ref{eq:upperdependence}) can be interpreted as that, if a large precipitation amount $\hat{p}_t$ is forecasted, there will be less room for the realistic amount to exceed the forecast value, which is in accordance with our intuition. Based on the above analysis, both the lower-bound and upper-bound of each dimension in $\boldsymbol{\xi}$, and hence the distribution of $\boldsymbol{\xi}$, depend on the forecast sequence $\hat{\bold{p}}$. From Fig. \ref{fig:Prediction_vs_Error}, the validity of (\ref{eq:lowerdependence}) and (\ref{eq:upperdependence}) can be clearly observed as well.

To derive an analytical expression of the uncertainty set for $\boldsymbol{\xi}$ that depends on $\hat{\bold{p}}$, we further assume that the uncertainty $\boldsymbol{\xi}$ can be realized in a \textit{generative} manner that $\boldsymbol{\xi}$ is a function of both external variable $\hat{\bold{p}}$ and the random variable $\bar{\boldsymbol{\xi}}$, i.e. $\boldsymbol{\xi} = f(\bar{\boldsymbol{\xi}},\hat{\bold{p}})$, where the distribution of $\bar{\boldsymbol{\xi}}$ is independent from $\hat{\bold{p}}$. In this way, existing uncertainty sets become applicable to describe the distribution of $\bar{\boldsymbol{\xi}}$, thereby indirectly characterizing the distribution of $\boldsymbol{\xi}$. Starting from these motivations, we describe the distribution of precipitation forecast errors $\boldsymbol{\xi}$ based on (\ref{eq:asymset}) by incorporating scaling information into matrices $\bold{C}$ and $\bold{D}$, which are functions of $\hat{\bold{p}}$:
\begin{gather}
\bold{C}(\hat{\bold{p}}) = {\rm diag}\{p^{\max} - \hat{p}_1,\cdots,p^{\max} - \hat{p}_H\}, \\
\bold{D}(\hat{\bold{p}}) = {\rm diag}\{\hat{p}_1,\cdots,\hat{p}_H\}.
\end{gather}
This turns out to parameterize the dependence of $\boldsymbol{\xi}$ on both $\hat{\bold{p}}$ and $\bar{\boldsymbol{\xi}}$:
\begin{equation}
\begin{split}
\boldsymbol{\xi} = &\ \bold{C}(\hat{\bold{p}})\boldsymbol{\xi}^+ - \bold{D}(\hat{\bold{p}})\boldsymbol{\xi}^- \\
= &\ \bold{C}(\hat{\bold{p}})\max\{\bar{\boldsymbol{\xi}}, \bold{0}\} - \bold{D}(\hat{\bold{p}})\max\{-\bar{\boldsymbol{\xi}}, \bold{0}\} \triangleq f(\bar{\boldsymbol{\xi}},\hat{\bold{p}})
\end{split}
\label{eq:asymset1}
\end{equation}
where $\max\{\bar{\boldsymbol{\xi}}, \bold{0}\} = \boldsymbol{\xi}^+ \le \bold{1}$ and $\max\{-\bar{\boldsymbol{\xi}}, \bold{0}\} = \boldsymbol{\xi}^- \le \bold{1}$ hold after normalizations, which in turn ensure the lower-bound (\ref{eq:lowerdependence}) and the upper-bound (\ref{eq:upperdependence}). In this sense, $\bar{\boldsymbol{\xi}}$ can be understood as the \textit{primitive uncertainty} that is independent of external variables $\hat{\bold{p}}$ and essentially induces randomness of $\boldsymbol{\xi}$. Because the distribution of $\bar{\boldsymbol{\xi}}$ may have an intrinsic complicated structure, we propose to use the SVC-based uncertainty set to excavate its support information from a set of data samples $\{ \bar{\boldsymbol{\xi}}^{(i)} \}_{i=1}^N$:
\begin{equation}
\bar{\boldsymbol{\xi}} = \boldsymbol{\xi}^+ - \boldsymbol{\xi}^- \in \mathcal{D}_{\bar{\boldsymbol{\xi}}},
\end{equation}
where
\begin{equation}
\mathcal{D}_{\bar{\boldsymbol{\xi}}} = \left \{ \bar{\boldsymbol{\xi}} \left | \sum\nolimits_{i \in {\rm SV}} \alpha_i \| \bold{Q}(\bar{\boldsymbol{\xi}} - \bar{\boldsymbol{\xi}}^{(i)}) \|_1 \le \theta \right . \right \}.
\label{eq:SVCset1}
\end{equation}
By summarizing the above results, we can arrive at the following expression of the \textit{conditional uncertainty set} to deliberately describe the distribution of precipitation forecast error $\boldsymbol{\xi}$:
\begin{equation}
\mathcal{D}_{\boldsymbol{\xi}}(\hat{\bold{p}}) = \left \{ \boldsymbol{\xi} \in \mathbb{R}^H \left | \begin{split}
& \boldsymbol{\xi} = \bold{C}(\hat{\bold{p}})\boldsymbol{\xi}^+ - \bold{D}(\hat{\bold{p}})\boldsymbol{\xi}^-\\
& \boldsymbol{\xi}^+ - \boldsymbol{\xi}^- \in \mathcal{D}_{\bar{\boldsymbol{\xi}}},\ \bold{0} \le \boldsymbol{\xi}^+, \boldsymbol{\xi}^- \le \bold{1}
\end{split}
\right . \right \}.
\label{eq:asym_set}
\end{equation}
Since $\mathcal{D}_{\bar{\boldsymbol{\xi}}}$ is a polytope, it can be easily deduced that $\mathcal{D}_{\boldsymbol{\xi}}(\hat{\bold{p}})$ is also a polytope.

We point out that the proposed conditional uncertainty set also applies to uncertainties that arise from predictions of \textit{some discrete events with bounded intensities}, e.g. cloud cover and building occupancy, provided that historical forecast and measurement data are available. In such cases, prediction errors also manifest asymmetric and dependent structures that are similar to precipitation prediction errors; hence, the proposed conditional uncertainty set enjoys generality in certain degrees.

Up to now, we can formulate the uncertainty set $\mathcal{W}$ of disturbance sequence $\bold{w}$ based on the Minkowski sum of two elementary uncertainty sets for $\boldsymbol{\eta}$ and $\boldsymbol{\xi}$. Because $\bold{w} = \boldsymbol{\xi} - \boldsymbol{\eta}$, we have $\bold{w} \in \mathcal{W} = \mathcal{D}_{\boldsymbol{\xi}}(\hat{\bold{p}}) + (-\mathcal{D}_{\boldsymbol{\eta}})$, which is obviously a polytope whose parameters depend on $\hat{\bold{p}}$ as well.

\section{Data-Driven Tuning of Uncertainty Sets and Performance Guarantees}
In this work, uncertainty sets are constructed in a data-driven manner. Therefore, their performance is of crucial importance in DDRMPC. A general criterion to certificate the performance of an uncertainty set $\mathcal{D}$ is called the $(1-\epsilon)$-prediction set \cite{hong16approximating}, which can be precisely expressed as $\mathbb{P} \{ \bold{w} \in \mathcal{D} \} \ge 1 - \epsilon$. However, because data sampling is prone to randomness, the constructed data-driven uncertainty set $\mathcal{D}$ itself is random. Therefore, we wish to make sure that the data-driven uncertainty set $\mathcal{D}$ is a $(1-\epsilon)$-prediction set with a suitably large confidence $1 - \beta$, which can be formally described as:
\begin{equation}
\mathbb{P}_\mathcal{D} \left \{ \mathbb{P} \left \{ \bold{w} \in \mathcal{D} \right \} \ge 1 - \epsilon \right \} \ge 1 - \beta,
\label{eq:probguarantee}
\end{equation}
where the outmost probability $\mathbb{P}_{\mathcal{D}} \{ \cdot \}$ refers to the probability of constructing the data-driven uncertainty set $\mathcal{D}$, and the innermost probability $\mathbb{P} \{ \cdot \}$ is taken respect to uncertainty $\bold{w}$ itself. $\epsilon$ and $\beta$ are user-specified parameters. Typically, one could set $\epsilon$ to be 0.05 or 0.10, and $\beta$ to be smaller than $10^{-3}$.

To attain the performance guarantee (\ref{eq:probguarantee}), we adopt the training-calibration procedure developed by \cite{hong16approximating} to further tune the size of the uncertainty set based on an independent set of calibration samples. It assumes that the uncertainty set can be parameterized as $\mathcal{D} = \{ \bold{w} | y(\bold{w}) \le \theta \}$, where $y(\cdot)$ is a scalar function; meanwhile, all available data samples have been split into a training dataset and a calibration dataset. After initially determining $y(\cdot)$ with the training dataset, the value of $\theta$ in $\mathcal{D}$ is then refined based on the calibration dataset $\{ \bold{w}^{(i)}_{\rm calib} \}_{i=1}^{N_{\rm calib}}$ of cardinality $N_{\rm calib}$ according to the following rule:
\begin{equation}
\theta \triangleq \max_{1 \le i \le N_{\rm calib}} \left \{ y\left (\bold{w}^{(i)}_{\rm calib}\right ) \right \}.
\label{eq:theta}
\end{equation}
If the number of calibration data satisfies $N_{\rm calib} \ge \log \beta / \log (1 - \epsilon)$, then the performance guarantee (\ref{eq:probguarantee}) can be secured for the uncertainty set $\mathcal{D}$ with the value of $\theta$ calibrated according to (\ref{eq:theta}) \cite{shang18data}. Therefore, this strategy can be directly applied to $\mathcal{D}_{\boldsymbol{\eta}}$, since the size of SVC-based uncertainty set in (\ref{eq:SVCset}) is governed by only one parameter $\theta$, and the scalar function $y(\cdot)$ corresponds to the LHS of the constraint in (\ref{eq:SVCset}). This yields the following performance guarantee for $\mathcal{D}_{\boldsymbol{\eta}}$:
\begin{equation}
\mathbb{P}_{\mathcal{D}_{\boldsymbol{\eta}}} \left \{ \mathbb{P} \left \{ \boldsymbol{\eta} \in \mathcal{D}_{\boldsymbol{\eta}}  \right \} \ge 1 - \epsilon \right \} \ge 1 - \beta.
\end{equation}

As for the conditional uncertainty set $\mathcal{D}_{\boldsymbol{\xi}}(\hat{\bold{p}})$ for $\boldsymbol{\xi}$, notice that we only need to ensure the same performance guarantee for $\mathcal{D}_{\bar{\boldsymbol{\xi}}}$, which is also an SVC-based uncertainty set. Therefore, the training-calibration strategy can be trivially applied. According to (\ref{eq:asymset1}) and (\ref{eq:asym_set}), if
\begin{equation}
\mathbb{P}_{\mathcal{D}_{\bar{\boldsymbol{\xi}}}} \left \{ \mathbb{P} \left \{ \bar{\boldsymbol{\xi}} \in \mathcal{D}_{\bar{\boldsymbol{\xi}}}  \right \} \ge 1 - \epsilon \right \} \ge 1 - \beta,
\end{equation}
is satisfied, it can be easily deduced that the performance guarantee
\begin{equation}
\mathbb{P}_{\mathcal{D}_{\boldsymbol{\xi}}(\hat{\bold{p}})} \left \{ \mathbb{P} \left \{ \boldsymbol{\xi} \in \mathcal{D}_{\boldsymbol{\xi}}(\hat{\bold{p}}) \right \} \ge 1 - \epsilon \right \} \ge 1 - \beta,
\end{equation}
will hold as well.

Since the uncertain disturbance sequence $\bold{w}$ in (\ref{eq:linearsystems}) is comprised of evapotranspiration forecast errors and precipitation forecast errors, which are described by two individual data-driven uncertainty sets, we are specifically interested in how well the distribution of $\bold{w}$ is described by $\mathcal{W}= \mathcal{D}_{\boldsymbol{\eta}} + (-\mathcal{D}_{\boldsymbol{\xi}}(\hat{\bold{p}}))$ in a probabilistic sense. Next, we provide the following theorem to establish appropriate performance guarantee for the Minkwoski sum of two data-driven uncertainty sets.

\textit{Theorem 1:} Let $\mathcal{D}_{\boldsymbol{\eta}}$ and $\mathcal{D}_{\boldsymbol{\xi}}(\hat{\bold{p}})$ be data-driven uncertainty sets of evapotranspiration forecast error $\boldsymbol{\eta}$ and precipitation forecast error $\boldsymbol{\xi}$, respectively. If $\mathcal{D}_{\boldsymbol{\eta}}$ and $\mathcal{D}_{\boldsymbol{\xi}}(\hat{\bold{p}})$ admit their respective performance guarantees:
\begin{gather}
\mathbb{P}_{\mathcal{D}_{\boldsymbol{\eta}}} \left \{ \mathbb{P} \left \{ \boldsymbol{\eta} \in \mathcal{D}_{\boldsymbol{\eta}}  \right \} \ge 1 - \epsilon_1 \right \} \ge 1 - \beta_1, \\
\mathbb{P}_{\mathcal{D}_{\boldsymbol{\xi}}(\hat{\bold{p}})} \left \{ \mathbb{P} \left \{ \boldsymbol{\xi} \in \mathcal{D}_{\boldsymbol{\xi}}(\hat{\bold{p}})   \right \} \ge 1 - \epsilon_2 \right \} \ge 1 - \beta_2,
\end{gather}
then the following performance guarantee for the data-driven uncertainty set $\mathcal{W}= \mathcal{D}_{\boldsymbol{\eta}} + (-\mathcal{D}_{\boldsymbol{\xi}}(\hat{\bold{p}}))$ holds:
\begin{equation}
\mathbb{P}_{\mathcal{W}} \left \{ \mathbb{P} \left \{ \bold{w} = \boldsymbol{\eta} - \boldsymbol{\xi} \in \mathcal{W} \right \} \ge 1 - \epsilon \right \} \ge 1 - \beta,
\label{eq:probguarantee1}
\end{equation}
where $\epsilon = \epsilon_1 + \epsilon_2,\ \beta = \beta_1 + \beta_2.$

\textit{Proof:} See the supplementary document.

Theorem 1 indicates that, it suffices to endow uncertainty sets $\mathcal{D}_{\boldsymbol{\eta}}$ and $\mathcal{D}_{\boldsymbol{\xi}}(\hat{\bold{p}})$ with performance guarantees separately, and then (\ref{eq:probguarantee}) will be automatically ensured. Given $\epsilon$ and $\beta$, one could trivially set $\epsilon_1 = \epsilon_2 = \epsilon / 2,\ \beta_1 = \beta_2 = \beta / 2$.

\section{Tractable Approximations with Generalized Affine Disturbance Feedback Law}
\subsection{Generalized Affine Disturbance Feedback}
Optimizing the infinite-dimensional closed-loop policy $\boldsymbol{\pi}(\bold{w})$ in the RMPC problem (\ref{eq:RMPC}) typically leads to an intractable problem. This motivates the use of the affine disturbance feedback (ADF) policy as a tractable approximation \cite{goulart06optimization,bental04adjustable,ning17data1}:
\begin{equation}
\begin{split}
\pi_t(\bold{w}) := \sum\nolimits_{j=0}^{t - 1} M_{t,j} (\xi_j - \eta_j) + h_t,\ t \in \mathbb{N}_{1:H}
\end{split}
\label{eq:adf}
\end{equation}
where $\pi_t(\bold{w})$ is restricted to depend linearly on all past disturbances. The ADF admits the following compact expression:
\begin{equation}
\boldsymbol{\pi}(\bold{w}) = \bold{M} \bold{w} + \bold{h} = \bold{M} (\boldsymbol{\xi} - \boldsymbol{\eta}) + \bold{h},
\label{eq:adf1}
\end{equation}
where
\begin{gather}
\bold{M} = \left [
\begin{array}{c c c c}
\bold{0} & \bold{0} & \cdots & \bold{0} \\
M_{1,0} & \bold{0} & \cdots & \bold{0} \\
\vdots & \vdots & \ddots & \vdots \\
M_{H,0} & M_{H,1} & \cdots & \bold{0}
\end{array}
\right ],\ \bold{h} = \left [
\begin{array}{c}
h_0 \\
h_{1} \\
\vdots \\
h_{H}
\end{array}
\right ].
\end{gather}
The causality of ADF policy is secured by the lower-triangular structure of $\bold{M}$. It has been proved in \cite{goulart06optimization} that ADF is identical to state feedback policy, which typically leads to non-convex optimization problems that are difficult to tackle online. In contrast, under ADF the optimal control problem (\ref{eq:RMPC}) of RMPC becomes the following optimization problem, with coefficients $\bold{M}$ and $\bold{h}$ in ADF being decision variables:
\begin{equation}
\begin{split}
\min_{\bold{M}, \bold{h}} & \ J_H(x, \bold{M}(\boldsymbol{\xi} - \boldsymbol{\eta}) + \bold{h}) \\
{\rm s.t.} &\ \bold{F}_x \left [ \bold{A}x_0 + (\bold{B}_u \bold{M} + \bold{B}_w)(\boldsymbol{\xi} - \boldsymbol{\eta}) + \bold{B}_u\bold{h} + \bold{B}_v \bold{v}  \right ] \le \bold{f}_x, \\
&\ \quad \quad \quad \quad \quad \quad \quad \quad \quad \quad \quad \quad \quad \quad \forall \boldsymbol{\eta} \in \mathcal{D}_{\boldsymbol{\eta}},\ \boldsymbol{\xi} \in \mathcal{D}_{\boldsymbol{\xi}}(\hat{\bold{p}}) \\
&\ \bold{F}_u \left [ \bold{M}(\boldsymbol{\xi} - \boldsymbol{\eta}) + \bold{h} \right ] \le \bold{f}_u,\ \forall \boldsymbol{\eta} \in \mathcal{D}_{\boldsymbol{\eta}},\ \boldsymbol{\xi} \in \mathcal{D}_{\boldsymbol{\xi}}(\hat{\bold{p}})
\end{split}
\label{eq:optproblem_adf}
\end{equation}
which can be further reformulated as a convex optimization problem using robust optimization techniques if $J_H(\cdot,\cdot)$ is convex \cite{goulart06optimization}. Despite of these merits, the ADF policy is inevitably prone to suboptimality. As a matter of fact, $\bold{w}$ can be expressed as a linear combination of the \textit{lifted uncertainties} $\boldsymbol{\xi}^+$, $\boldsymbol{\xi}^-$ and $\boldsymbol{\eta}$:
\begin{equation}
\bold{w} = \boldsymbol{\xi} - \boldsymbol{\eta} = \bold{C}(\hat{\bold{p}})\boldsymbol{\xi}^+ - \bold{D}(\hat{\bold{p}})\boldsymbol{\xi}^- - \boldsymbol{\eta}.
\end{equation}
Then we can make further improvement by enforcing affine dependence of $\boldsymbol{\pi}(\cdot)$ directly on $\boldsymbol{\xi}^+$, $\boldsymbol{\xi}^-$ and $\boldsymbol{\eta}$:
\begin{equation}
\pi_t(\bold{w}) := \sum\nolimits_{j=0}^{t - 1} \left ( M_{t,j}^+ \xi_{j}^+ + M_{t,j}^- \xi_{j}^- + L_{t,j}\eta_j \right ) + h_t,
\end{equation}
which is termed as \textit{generalized affine decision rule} (GADF) based on lifted uncertainties \cite{georghiou15generalized,zhang17robust}. The GADF also allows for a compact expression:
\begin{equation}
\boldsymbol{\pi}(\bold{w}) = \bold{M}^+ \boldsymbol{\xi}^+ + \bold{M}^- \boldsymbol{\xi}^- + \bold{L} \boldsymbol{\eta} + \bold{h},
\end{equation}
where $\bold{M}^+$, $\bold{M}^-$ and $\bold{L}$ are lower-triangular matrices with all diagonal elements being zeros, thereby ensuring the causality of GADF. By imposing dependence of GADF on lifted uncertainties, the conservatism of approximations can be further reduced \cite{georghiou15generalized,zhang17robust}. For this reason, the GADF policy is adopted in this work, which yields the following optimization problem:
\begin{equation}
\begin{split}
\min_{\begin{subarray}{c}
\bold{h}, \bold{M}^+ \\
\bold{M}^-, \bold{L}
\end{subarray}} & \ J_H(x, \bold{M}^+ \boldsymbol{\xi}^+ + \bold{M}^- \boldsymbol{\xi}^- + \bold{L} \boldsymbol{\eta} + \bold{h}) \\
{\rm s.t.}\ &\ \bold{F}_x \left [ \bold{A}x_0 + (\bold{B}_u\bold{M}^+ + \bold{B}_w\bold{C}(\hat{\bold{p}}))\boldsymbol{\xi}^+ + (\bold{B}_u\bold{L} - \bold{B}_w)\boldsymbol{\eta} \right . \\
&\ \quad \ \ + (\bold{B}_u\bold{M}^- - \bold{B}_w\bold{D}(\hat{\bold{p}}))\boldsymbol{\xi}^- + \left . \bold{B}_u \bold{h} + \bold{B}_v \bold{v} \right ] \le \bold{f}_x, \\
&\ \quad \quad \quad \quad \forall \boldsymbol{\eta} \in \mathcal{D}_{\boldsymbol{\eta}},\ \boldsymbol{\xi}^+ - \boldsymbol{\xi}^- \in \mathcal{D}_{\bar{\boldsymbol{\xi}}},\ \bold{0} \le \boldsymbol{\xi}^+, \boldsymbol{\xi}^- \le \bold{1} \\
&\ \bold{F}_u \left ( \bold{M}^+ \boldsymbol{\xi}^+ + \bold{M}^- \boldsymbol{\xi}^- + \bold{L} \boldsymbol{\eta} + \bold{h} \right ) \le \bold{f}_u,\\
&\ \quad \quad \quad \quad \forall \boldsymbol{\eta} \in \mathcal{D}_{\boldsymbol{\eta}},\ \boldsymbol{\xi}^+ - \boldsymbol{\xi}^- \in \mathcal{D}_{\bar{\boldsymbol{\xi}}},\ \bold{0} \le \boldsymbol{\xi}^+, \boldsymbol{\xi}^- \le \bold{1} \\
\end{split}
\label{eq:optproblem}
\end{equation}
where $\bold{M}^+,\bold{M}^-,\bold{L},\bold{h}$ become decision variables to be optimized. The advantage of GADF over ADF can be formally demonstrated in terms of the following theorem.

\textit{Theorem 2:} For every possible feasible solution $\{ \bold{M},\bold{h} \}$ of problem (\ref{eq:optproblem_adf}), there always exists a solution $\{ \bold{M}^+, \bold{M}^-, \bold{L}, \bold{h} \}$ that is feasible for problem (\ref{eq:optproblem}) and has the same objective value.

The proof can be made in a similar spirit to that of Proposition 3 in \cite{zhang17robust}, and is hence omitted here for brevity. Theorem 2 indicates that solving the GADF-induced problem (\ref{eq:optproblem}) yields at least as good results as solving the classic ADF-induced problem (\ref{eq:optproblem_adf}). In this way, the conservatism brought by ADF approximation can be possibly reduced.

\subsection{Tractable Reformulations of Robust Constraints}
Next, we provide results to deal with the tractability of robust state and input constraints in the problem (\ref{eq:optproblem}). Since the left hand side (LHS) of each constraints is linear in the lifted uncertainties $\boldsymbol{\xi}^+$, $\boldsymbol{\xi}^-$ and $\boldsymbol{\eta}$, it can be regarded as a worst-case maximization problem with the objective being linear in $\boldsymbol{\xi}^+$, $\boldsymbol{\xi}^-$ and $\boldsymbol{\eta}$. Therefore, we only need to transform the associated worst-case optimization problems into their equivalent minimization problems, which are also referred to as robust counterparts, thereby ensuring the tractability of (\ref{eq:optproblem}).

Because $\boldsymbol{\eta}$ is decoupled from $\boldsymbol{\xi}^+$ and $\boldsymbol{\xi}^-$ in the LHS of all input and state constraints in (\ref{eq:optproblem}), we first deal with worst-case problems related to $\boldsymbol{\eta}$. To this end, we adopt the following lemma from literature, which is in fact the tractability result of the SVC-based uncertainty set.

\textit{Lemma 1 \cite{shang17data}:} The worst-case performance on the SVC-based uncertainty set, which is the optimal value of the following problem
\begin{equation}
\begin{split}
\max_{\boldsymbol{\eta}} &\ \bold{a}^{\rm T} \boldsymbol{\eta} \\
{\rm s.t.}\ & \sum\nolimits_{i \in {\rm SV}} \alpha_i \| \bold{Q} (\boldsymbol{\eta} - \boldsymbol{\eta}^{(i)}) \|_1 \le \theta
\end{split}
\end{equation}
is equal to the optimal value of the following linear program (LP):
\begin{equation}
\begin{split}
\min_{\boldsymbol{\mu}_i,\boldsymbol{\lambda}_i,k} &\ \sum\nolimits_{i \in {\rm SV}} (\boldsymbol{\mu}_i - \boldsymbol{\lambda}_i)^{\rm T} \bold{Q} \boldsymbol{\eta}^{(i)} + k\theta  \\
{\rm s.t.}\ &\ \sum\nolimits_{i \in {\rm SV}} \bold{Q}(\boldsymbol{\lambda}_i - \boldsymbol{\mu}_i) + \bold{a} = \bold{0} \\
&\ \boldsymbol{\lambda}_i + \boldsymbol{\mu}_i = k \cdot \alpha_i \cdot \bold{1},\ \forall i \in {\rm SV} \\
&\ \boldsymbol{\lambda}_i, \boldsymbol{\mu}_i \ge \bold{0},\ k \ge 0
\end{split}
\end{equation}

As with the lifted uncertainties $\boldsymbol{\xi}^+$ and $\boldsymbol{\xi}^-$, we establish the following theorem in this paper to address the tractability issue.

\textit{Theorem 3:} The optimal value of the worst-case maximization problem
\begin{equation}
\begin{split}
\max_{\boldsymbol{\xi}^+, \boldsymbol{\xi}^-} &\ \bold{a}^{\rm T} \boldsymbol{\xi}^+ + \bold{b}^{\rm T} \boldsymbol{\xi}^- \\
{\rm s.t.}\ & \sum\nolimits_{i \in {\rm SV}}\alpha_i \|\bold{Q}( \boldsymbol{\xi}^+ - \boldsymbol{\xi}^- - \bar{\boldsymbol{\xi}}^{(i)} )\|_1 \le \theta \\
            & \bold{0} \le \boldsymbol{\xi}^+,\boldsymbol{\xi}^- \le \bold{1}
\end{split}
\label{eq:worstcase}
\end{equation}
is equal to that of the following LP:
\begin{equation}
\begin{split}
\min_{\boldsymbol{\mu}_i,\boldsymbol{\lambda}_i,\bold{r},\bold{s},k} &\ \sum\nolimits_{i \in {\rm SV}} (\boldsymbol{\mu}_i - \boldsymbol{\lambda}_i)^{\rm T} \bold{Q} \bar{\boldsymbol{\xi}}^{(i)} + (\bold{r} + \bold{s})^{\rm T}\bold{1} + k\theta \\
{\rm s.t.}\ \ \ &\ \sum\nolimits_{i \in {\rm SV}} \bold{Q}(\boldsymbol{\mu}_i - \boldsymbol{\lambda}_i) + \bold{r} \ge \bold{a} \\
&\ \sum\nolimits_{i \in {\rm SV}} \bold{Q}(\boldsymbol{\lambda}_i - \boldsymbol{\mu}_i) + \bold{s} \ge \bold{b} \\
&\ \boldsymbol{\lambda}_i + \boldsymbol{\mu}_i = k \cdot \alpha_i \cdot \bold{1},\ \forall i \in {\rm SV} \\
&\ \boldsymbol{\lambda}_i, \boldsymbol{\mu}_i, \bold{r}, \bold{s} \ge \bold{0},\ k \ge 0
\end{split}
\label{eq:worstcase_dual}
\end{equation}

\textit{Proof:} See the supplementary document.

Therefore, both input and state constraints in (\ref{eq:optproblem}) can be recast as a series of linear inequalities and equalities, with additional decision variables involved. Finally, if the objective $J_H(x,\boldsymbol{\pi})$ is convex in $\boldsymbol{\pi}$, we are able to reformulate (\ref{eq:optproblem}) as a convex optimization problem that can be readily solved in practice, thereby ensuring the practicability of DDRMPC.

\section{Simulated Case Studies}
\subsection{Problem Description}
In this section, we carry out closed-loop simulation case studies based on real weather condition data collected at Des Moines, Iowa, US to validate performances of different control methods. In the dynamic model (\ref{eq:state_space}), the sampling interval is set as 6 h, and the value of $c$ is determined as 0.025. Historical weather condition data from May 2017 to Oct. 2017 have been collected from \cite{iowa}, based on which closed-loop simulations are performed.

In this work, our control goal is to maintain the soil moisture above a safety level $x_{\min} = 30$ mm with minimum water usage under the maximum water supply $u_{\max} = 10$ mm. More general goals can also be designed to achieve efficient control of crop yield, growth rate, disease and quality. We employ two MPC techniques, i.e. the proposed DDRMPC and the certainty equivalent MPC (CEMPC), where the prediction horizon of the dynamic model (\ref{eq:linearsystems}) has $H = 8$ intervals. To minimize water usage, we use the expected cost function (\ref{eq:expectedcost}) with cost-to-go function $l(x,u) = u^{\rm T}u$ and terminal cost $l_f(x) = 0$, and various constraints can be systematically incorporated in the MPC framework. To optimize the system behavior in the near future, forecasts of both evapotranspiration and precipitation are needed, while in CEMPC, it is assumed that all forecasts are accurate. We collect 48-hour ahead forecast data for 6-hour cumulative precipitations $\hat{\bold{p}}$ from \cite{iowa}. As for evapotranspiration forecast $\hat{\bold{e}}$, we adopt the simple prediction model established in \cite{hargreaves85reference}:
\begin{equation}
\hat{e}_t = \Gamma_c \times {\rm RA} \times \sqrt{{\rm TD}} \left ( T_t + 17.8^{\rm o}{\rm C} \right ),
\label{eq:evap}
\end{equation}
where $\Gamma_c$ is a crop-specific parameter. In this work, the Alta fescue grass is considered and the value of $\Gamma_c$ is set as $0.0023$ accordingly \cite{hargreaves85reference}. RA stands for the extraterrestrial radiation, which is in the same unit as $\hat{e}_t$. TD denotes the annual average daily temperature difference, which can be derived from local meteorological data, and $T_t$ is the average outdoor temperature during the $t$th time period. Since outdoor temperature forecasts for $T_t$ are also available from \cite{iowa}, predictions of future evapotranspiration $\hat{\bold{e}}$ can be readily obtained based on (\ref{eq:evap}). Note that the proposed DDRMPC framework is compatible not only with the evapotranspiration forecast model (\ref{eq:evap}) but also with more general models, because the uncertainty is described in a data-driven manner. Here the forecast model (\ref{eq:evap}) merely serves as a simple example, and in practice one can use more flexible models based on other micro-meteorological and environmental factors such as cloud cover and relative humidity to improve the accuracy.

To establish two data-driven uncertainty sets in DDRMPC, we collect both forecast data and measurement data from May 2016 to Oct. 2016, which yields $729$ scenarios in total for forecast errors $\boldsymbol{\eta}$ and $\boldsymbol{\xi}$. We set $\epsilon = 0.05$ and $\beta = 10^{-4}$, which indicate that with $99.99\%$ confidence, the data-driven uncertainty set covers $95\%$ of probability mass of uncertainty. It leads to $\epsilon_1 = \epsilon_2 = 0.025$ and $\beta_1 = \beta_2 = 5 \times 10^{-5}$, and hence $N_{\rm calib} = 387$ samples are used as calibration data. The rest $342$ data samples are used for training SVC models and deriving initial data-driven uncertainty sets. In the conditional uncertainty set $\mathcal{D}_{\boldsymbol{\xi}}(\hat{\bold{p}})$, the maximum 6-hour cumulative precipitation amount in each time period is set as $p^{\max} = 50$ mm. It is worth mentioning that, user-specified parameters such as $p^{\max}$, $x_{\min}$ and $u_{\max}$ can also be adjusted flexibly in order to account for time-varying seasonal factors.

Finally, optimal control problems of both proposed DDRMPC and CEMPC can be cast as QPs. The numerical solver {\tt cvx} in MATLAB \cite{grant08matlab} is adopted and all computations are carried out on a desktop computer with an Intel Core i7-6700 processor at 3.40 GHz and 32 GB of RAM. After deriving the optimal solution, only the first control decision is implemented onto the irrigation system, and this procedure is repeated in a receding horizon manner.

\subsection{Setups for Generic Control Strategies}
In addition to two MPC techniques, two classic irrigation control strategies, i.e. the simple open-loop control \cite{jensen80design} and the rule-based control \cite{goodwin95irrigation}, are also implemented in the simulations for a comprehensive comparison. In the open-loop control, the irrigation amount is scheduled weekly according to the criterion $u_{t+\tau} \equiv \max \{ b - ax_t, 0 \},\ \forall \tau \in \mathbb{N}_{0:27}$. It indicates that the irrigation amount in the upcoming week ($7 \times 4 = 28$ time periods in total) is pre-determined based on the soil moisture level at the beginning, and the lower the soil moisture level $x_t$, the larger the irrigation amount $u_{t+\tau}\ (0\le\tau\le27)$. Coefficients $a$ and $b$ are fine-tuned to minimize total water usage while maintaining zero probability of constraint violations, and the results under different coefficient values are shown in Fig. \ref{fig:openloop}. Accordingly, coefficients are chosen as $a = 0.07$ and $b = 6.4$ mm.
\begin{figure}[htb]
\centering
\includegraphics[width = 0.4\textwidth]{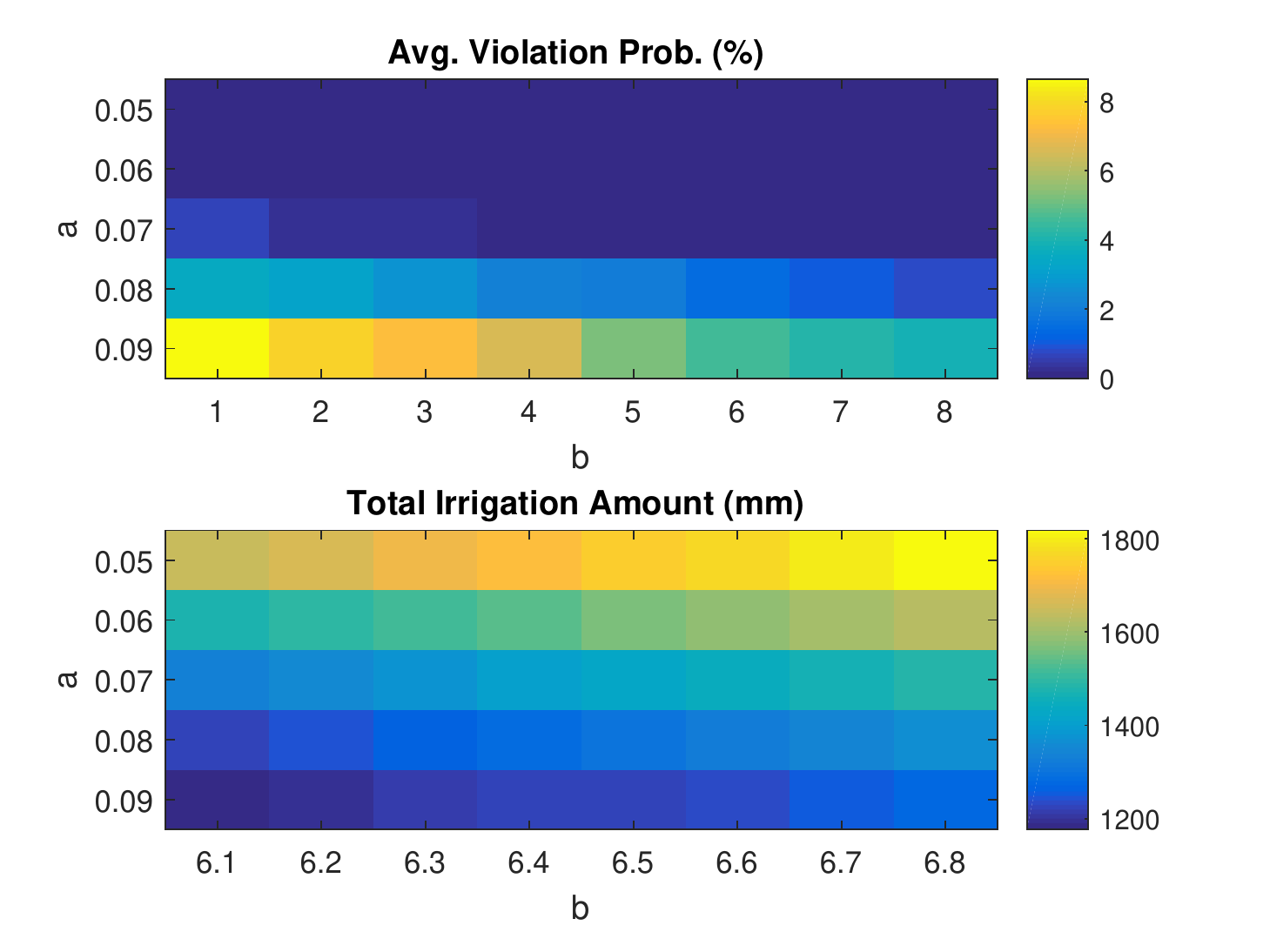}
\caption{Heat maps of average violation probability and total irrigation amount in the open-loop control.}
\label{fig:openloop}
\end{figure}

In the rule-based control, after the soil moisture level is detected to fall below a certain threshold $\delta$, a constant amount of water $C$ will be supplied, that is,
\begin{equation}
u_t = \left \{
\begin{array}{c c}
C, & {\rm if}\ x_t \le \delta \\
0, & {\rm if}\ x_t > \delta \\
\end{array}
\right .
\end{equation}
Control performances induced by different parameters values of $\{\delta,C\}$ are reported in Fig. \ref{fig:openloop}. We choose parameters that can safely maintain soil moisture above $30$ mm with minimum water usage. According to Fig. \ref{fig:openloop}, the threshold $\delta$ and the irrigation amount $C$ is determined as $33$ mm and $3$ mm, respectively.
\begin{figure}[htb]
\centering
\includegraphics[width = 0.4\textwidth]{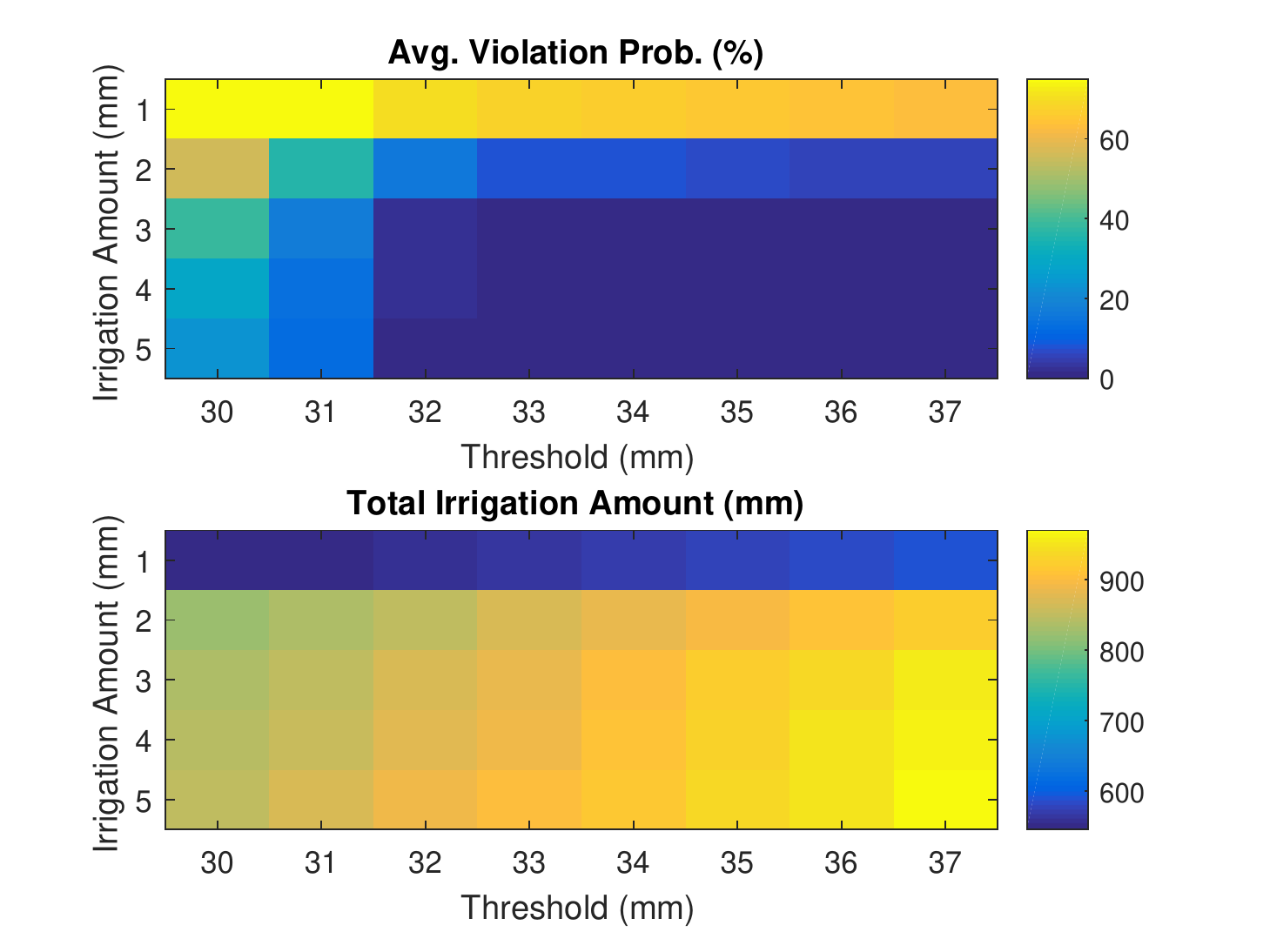}
\caption{Heat maps of average violation probability and total irrigation amount in the rule-based control.}
\label{fig:openloop}
\end{figure}

\subsection{Results and Discussions}
\begin{figure*}[t]
\centering
\includegraphics[width = 0.7\textwidth]{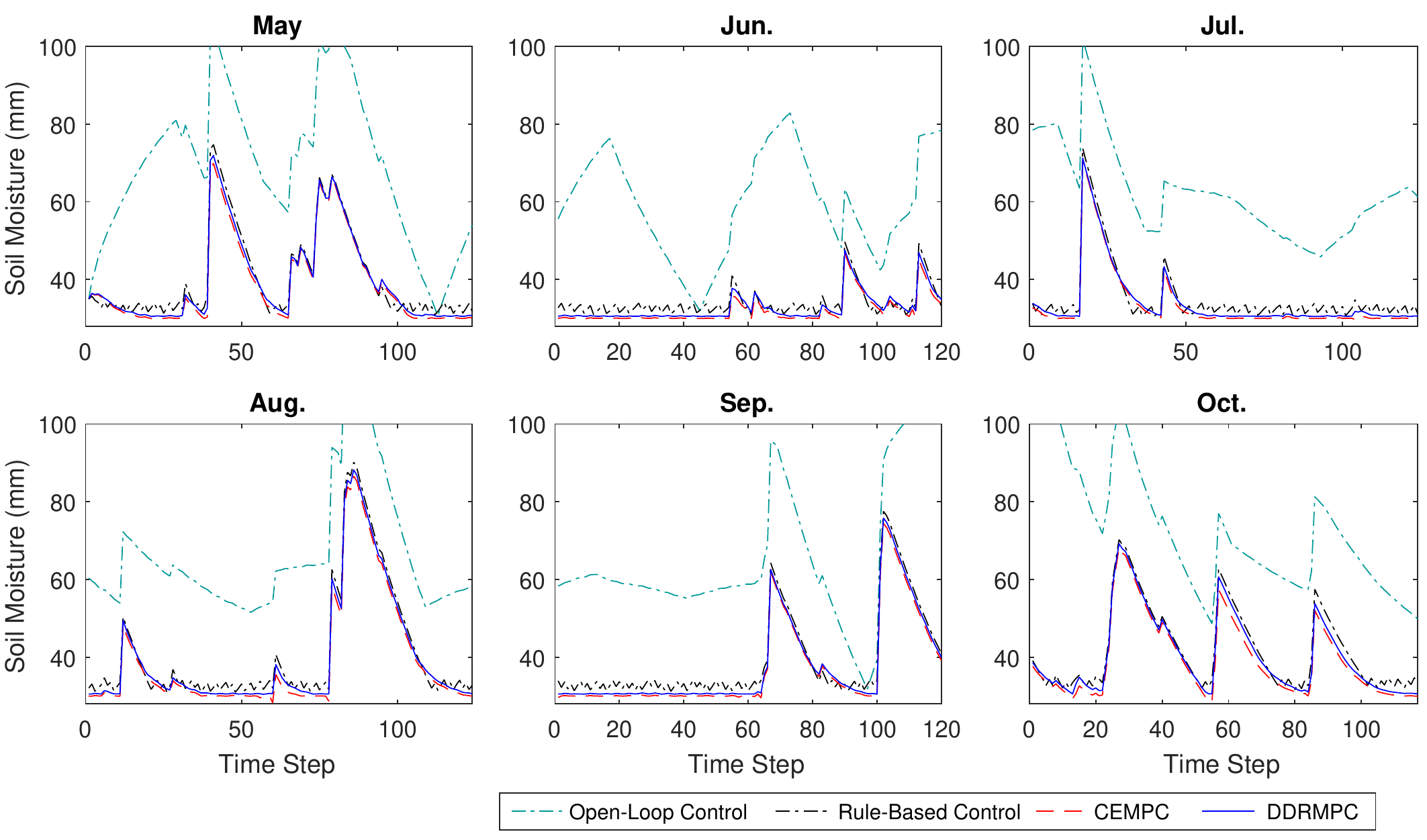}
\caption{Soil moisture levels under different control strategies.}
\label{fig:x}
\end{figure*}

We report control performances of simulations from May 2017 to Oct. 2017 under four different control strategies, namely the open-loop control, the rule-based control, CEMPC and DDRMPC. In Fig. \ref{fig:x}, state profiles of soil moisture levels are plotted, where the soil moisture lies in a low level most of the time, and increases abruptly when there is heavy precipitation. However, after the precipitation ends, the soil moisture level tends to decrease due to the effect of runoff and deep percolation. We can observe that the open-loop control yields the worst performance, whereas the proposed DDRMPC and CEMPC can operate the system much closer to the safety level (30 mm) than the rule-based control. It indicates that it is necessary to apply closed-loop irrigation control methods, and further incorporating meaningful information of weather forecast in closed-loop control is helpful for improving performance. In addition, CEMPC induces the least conservative control decisions. This is also justified by statistics with respect to monthly irrigation amounts, which are reported in Table \ref{tab:irrigation}. This is reasonable because CEMPC assumes forecasts to be precise and provides no allowance for robustness. Therefore, when weather forecast predicts the possibility of precipitation in the near future, CEMPC provides less water supply confidently with the aim to avoid unnecessary water usage.
\begin{table*}[htb]
\caption{Irrigation Amounts in 2017}
\centering
\begin{tabular}{c c c c c c c c}
\hline
     & May & Jun. & Jul. & Aug. & Sep. & Oct. & Total \\
\hline
Open-Loop Control (mm) & 234.00 & 300.00 & 283.00 & 226.00 & 268.00 & 92.00 & 1403.00 \\
Rule-Based Control (mm) & 123.00 & 201.00 & 216.00 & 150.00 & 135.00 & 57.00 & 882.00 \\
CEMPC (mm) & 115.20 & 196.64 & 209.03 & 141.11 & 131.12 & 47.32 & 840.42 \\
SP Tracking MPC (mm) & 121.96 & 199.47 & 219.24 & 146.14 & 134.44 & 52.51 & 873.75 \\
DDRMPC (mm) & 118.03 & 199.46 & 210.62 & 144.42 & 133.12 & 50.63 & 856.28 \\
DDRMPC under ADF Policy (mm) & 118.17 & 199.67 & 210.64 & 144.55 & 133.24 & 50.78 & 857.05 \\
\hline
\end{tabular}
\label{tab:irrigation}
\end{table*}

\begin{table*}[htb]
\caption{Water Loss Due to Runoff and Percolation Amounts in 2017}
\centering
\begin{tabular}{c c c c c c c c}
\hline
     & May & Jun. & Jul. & Aug. & Sep. & Oct. & Total \\
\hline
Open-Loop Control (mm) & 213.64 & 178.94 & 193.07 & 207.02 & 198.32 & 213.59 & 1204.58 \\
Rule-Based Control (mm) & 125.90 & 101.95 & 109.25 & 127.51 & 118.14 & 122.25 & 705.00 \\
CEMPC (mm) & 120.81 & 95.96 & 103.29 & 120.80 & 111.47 & 114.49 & 666.82 \\
SP Tracking MPC (mm) & 124.61 & 101.76 & 110.54 & 126.12 & 116.97 & 117.57 & 697.57 \\
DDRMPC (mm) & 123.10 & 98.28 & 105.37 & 123.76 & 113.42 & 118.21 & 682.14 \\
DDRMPC under ADF Policy (mm) & 123.25 & 98.39 & 105.49 & 123.88 & 113.49 & 118.37 & 682.87 \\
\hline
\end{tabular}
\label{tab:runoff}
\end{table*}

Table \ref{tab:runoff} further shows water losses due to runoff and percolation of different control strategies, where CEMPC leads to the least water loss in each month. This is because CEMPC maintains the soil moisture at a lower level, thereby effectively reducing the effect of runoff and percolation, which is proportional to the soil moisture level. Nevertheless, a fatal limitation of CEMPC is that, it fails to provide sufficient protection for constraints on soil moisture levels. In Table \ref{tab:violation}, probabilities of state constraint violations in each month are calculated and reported. It can be seen that the proposed DDRMPC and rule-based control strategy can always maintain the soil moisture level above 30 mm, whereas constraint violations frequently occur for CEMPC, which are particularly harmful for crop quality and productivity, and even result in a complete loss of crops. When heavy precipitation is predicted, CEMPC reduces the irrigation amount with the aim to fully utilize future precipitations. However, if the real precipitation amount turns out to be smaller than its forecast value, soil moisture deficiency will occur.
\begin{table*}[htb]
\caption{Probabilities of Constraint Violations in 2017}
\centering
\begin{tabular}{c c c c c c c c}
\hline
     & May & Jun. & Jul. & Aug. & Sep. & Oct. & Average \\
\hline
Open-Loop Control (\%) & 0 & 0 & 0 & 0 & 0 & 0 & 0 \\
Rule-Based Control (\%) & 0 & 0 & 0 & 0 & 0 & 0 & 0 \\
CEMPC (\%) & 8.06 & 21.67 & 23.38 & 15.32 & 17.50 & 5.98 & 15.36 \\
SP Tracking MPC (\%) & 0 & 0 & 0 & 0 & 0 & 0.85 & 0.14 \\
DDRMPC (\%) & 0 & 0 & 0 & 0 & 0 & 0 & 0 \\
DDRMPC under ADF Policy (\%) & 0 & 0 & 0 & 0 & 0 & 0 & 0 \\
\hline
\end{tabular}
\label{tab:violation}
\end{table*}

From these comparisons, the merits of the proposed DDRMPC can be explained as that, by extracting useful information from historical uncertainty data, a desirable balance is made between utilizing precipitation for reducing water usage and hedging against uncertainty in the forecast error. Although the open-loop control and the rule-based control achieve zero probabilities of moisture deficit, a large water consumption is inevitable. Compared to open-loop control, we can save $40\%$ of water consumption by utilizing forecast information in the control horizon, while still maintaining soil moisture above the safety levels. In comparison with closed-loop control strategies, improved control performances can also be obtained by DDRMPC, mainly because DDRMPC desirably provides a certain degree of robustness. Although DDRMPC consumes slightly more water than CEMPC, it eventually pays off because a satisfactory performance is achieved in safeguarding soil moisture level constraints under disturbances, which is particularly important to ensure crop quality and yield. In this sense, by using data information, the proposed DDRMPC can achieve a desirable tradeoff between using forecast information for saving water, and hedging against uncertainty in forecast error. In general, poorer forecast accuracy leads to larger uncertainty in forecast errors. In response to this increased uncertainty, DDRMPC provides more conservative decisions to ensure tight constraints that significantly influence crop quality and yield.

\begin{figure}[htb]
\centering
\includegraphics[width = 0.4\textwidth]{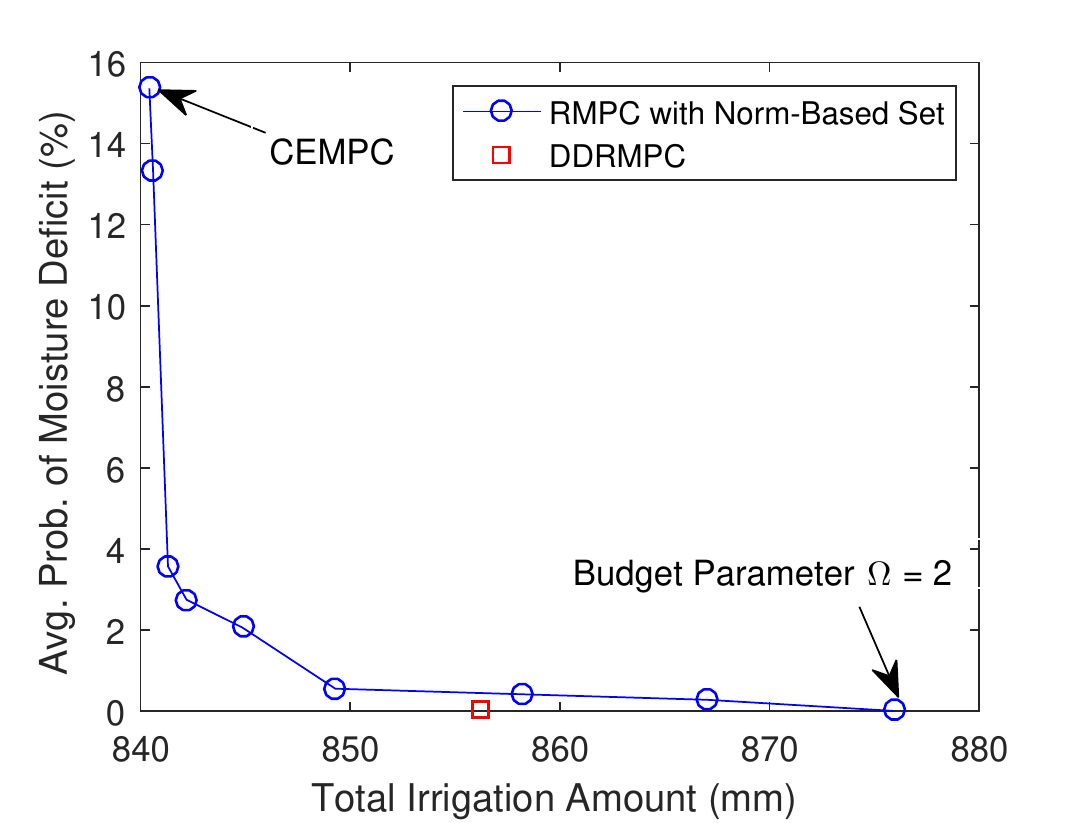}
\caption{Overall control performance comparison between classic RMPC with norm-based sets and DDRMPC in 2017.}
\label{fig:RMPC}
\end{figure}

Next, we investigate control performances of classic RMPC strategies, where the generic norm-based uncertainty set
\begin{equation}
\bold{w} \in \mathcal{W} = \left \{ \bold{w} \left | \| \bold{w} \|_1 \le \Omega \right . \right \},
\label{eq:normset}
\end{equation}
and the ADF policy are adopted in the optimal control problem (\ref{eq:RMPC}). Here different budget parameters $\Omega$ are used in the uncertainty set (\ref{eq:normset}), that is, $\Omega \in \{0, 0.01, 0.05, 0.1, 0.25, 0.5, 1, 1.5, 2 \}$, thereby giving rise to a variety of tradeoffs between efficiency and robustness. Hence, the overall control performances in terms of average probability of constraint violations and total irrigation amount in 2017 are plotted and compared against DDRMPC in Fig. \ref{fig:RMPC}. Note that when $\Omega = 0$, RMPC will reduce to CEMPC, which assumes perfect forecasts and hence brings frequent constraint violations. With the value of $\Omega$ increasing, the size of $\mathcal{W}$ becomes larger, which brings more robustness and hence continually reduces the probability of moisture deficiencies at the price of more water to be consumed for irrigation. However, its performance is much worse than that of DDRMPC. Fig. \ref{fig:RMPC} indicates that in order to ensure zero probability of constraint violations, we need to choose $\Omega = 2$, resulting in the total irrigation amount of 875.94 mm, which is higher than DDRMPC. This is because in generic RMPC, the uncertainty set with a fixed shape is adopted to delineate the support of uncertainties, thereby falling short of capturing the dependence of distribution on forecast values and leading to over-conservative solutions. For example, when the forecast states no precipitation in the near future, the worst-case uncertainty $\bold{w}$ shall have a small magnitude. In this case, a fixed uncertainty set inevitably incorporates excessive large uncertainty and leads to over-conservatism. This clearly demonstrates that by actively learning uncertainty information from data and elaborately designing uncertainty sets for various forecast errors, the conservatism of RMPC can be considerably reduced and hence better control performance can be obtained by DDRMPC, thereby demonstrating the value of data in control design.

We provide more details to comprehensively illustrate the practicability of implementing the proposed DDRMPC framework. The computational complexities of solving optimization problems in various MPC methods are reported in Table \ref{tab:complexity}. Compared with CEMPC and RMPC with $\Omega = 2$, the optimal control problem is more complicated in DDRMPC because a large number of Lagrange multipliers have been introduced to transform the infinite-dimensional problem into its robust counterpart. However, the computation is still thrifty since all constraints can still be cast as linear inequalities and equalities in the light of GADF, which can be readily handled by convex programming techniques. The average CPU time for DDRMPC is only 1.18s, which is acceptable in practice since it suffices to finish the optimization procedure within the sampling interval of 6 h.
\begin{table}[htb]
\caption{Computational Complexities of Different MPC Approaches}
\centering
\begin{tabular}{c c c c}
\hline
     & CEMPC & DDRMPC & RMPC ($\Omega = 2$) \\
\hline
\# of Variables & 27 & 4,247 & 147 \\
\# of Constraints & 9 & 2,137 & 45 \\
Avg. CPU Time (s) & 0.26 & 1.18 & 0.36 \\
\hline
\end{tabular}
\label{tab:complexity}
\end{table}

We also examined the control performance of a carefully tuned set-point (SP) tracking MPC, with the SP chosen as 33 mm. The resulting statistics are reported in Tables \ref{tab:irrigation}, \ref{tab:runoff}, and \ref{tab:violation}. Although its performance is better than that of rule-based control, with nearly zero violation probability, it still consumes more water than DDRMPC. Meanwhile, to seek a good performance and avoid constraint violations, one needs to carefully specify the SP as a key parameter. However, the relation between the SP and the violation probability is not clear. By contrast, DDRMPC yields a more convenient data-driven way to adjust parameters ($\epsilon$ and $\beta$) and control the conservatism.

Finally, we make a further comparative study to uncover the advantages of using GADF as the approximated control law. For comparison, the ADF policy (\ref{eq:adf}) is adopted, and we can still formulate the robust optimal control problem and translate it into an equivalent tractable robust counterpart. Under control law obtained in this way, closed-loop simulations have been carried out, and the results are reported in Tables \ref{tab:irrigation}-\ref{tab:violation} as well. It can be seen that under the ADF policy, a slightly worse control performance is obtained due to heavier water usage and increased water loss. This is because the suboptimal solution obtained by ADF approximation is always not better than that obtained by GADF approximation, thereby revealing the advantages of GADF policies.

\section{Conclusions and Recommendations}
In this paper, we develop a novel DDRMPC framework for efficient irrigation control. To optimize trajectory of future soil moisture levels and minimize water usage, predictions of evapotranspiration and precipitation are considered in the RMPC framework. To better capture the support of uncertainty distribution, we use two data-driven uncertainty sets to characterize possible realizations of evapotranspiration and precipitation forecast errors. For evapotranspiration forecast error, the SVC-based uncertainty set is adopted, which can be conveniently built from historical data. Because precipitation is a rare event with bounded intensities, we proposed the notion of conditional uncertainty set to describe the dependence of distributions of forecast errors on forecast values. To endow uncertainty sets with appropriate probabilistic guarantees, a practical training-calibration procedure is employed. GADF is adopted to approximately and efficiently solve the induced optimization problems in DDRMPC. The closed-loop simulation results demonstrate that, by actively learning uncertainty information from historical data, DDRMPC can reliably maintain soil moisture above the safety level and avoid crop devastation, with total water consumption reduced by $40\%$ compared to the fine-tuned open-loop control strategy. It also achieves significant improvements in control performance in comparison with classic carefully tuned rule-based control and certainty equivalent model predictive control.

In future work, more efforts could be made in incorporating more economic and environmental indices into control objectives and constraints. Another direction is to investigate tree crops of higher values, where the central focus is placed on the crop quality rather than the yield. The key issue arises from the complicated mechanistic models of tree crops, where biological and physical mechanisms are highly nonlinear and hence pose challenges in real-time optimizations.

\ifCLASSOPTIONcaptionsoff
  \newpage
\fi

\vspace{-2cm}

\begin{IEEEbiography}[{\includegraphics[width=1in,height=1.25in,clip,keepaspectratio]{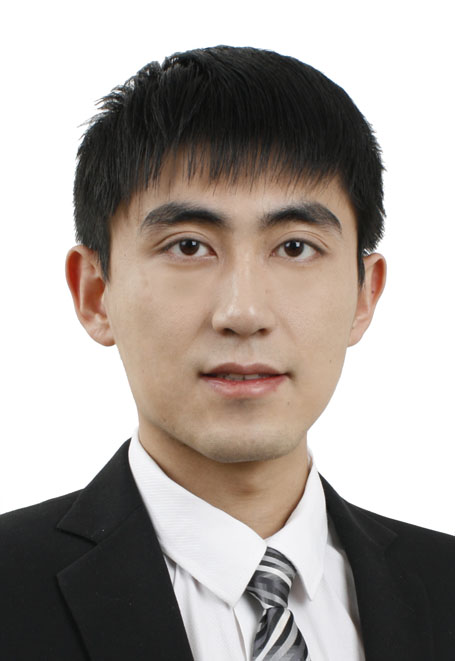}}]{Chao Shang} (M'17) received the B.Eng. degree in automation and the Ph.D. degree in control science and engineering from Tsinghua University, Beijing, China, in 2011 and 2016, respectively. After working as a Postdoctoral Fellow at Cornell University, he joined the Department of Automation, Tsinghua University in 2018 as an Assistant Professor. His research interests include data-driven modeling, monitoring, diagnosis and optimization with applications to industrial manufacturing processes.

Dr. Shang received the Springer Excellent Doctorate Theses Award in 2018, the Zijing Scholarship, the Outstanding Graduate Award and the Teaching Achievement Award from Tsinghua University in 2016.
\end{IEEEbiography}

\vspace{-2cm}

\begin{IEEEbiography}[{\includegraphics[width=1in,height=1.25in,clip,keepaspectratio]{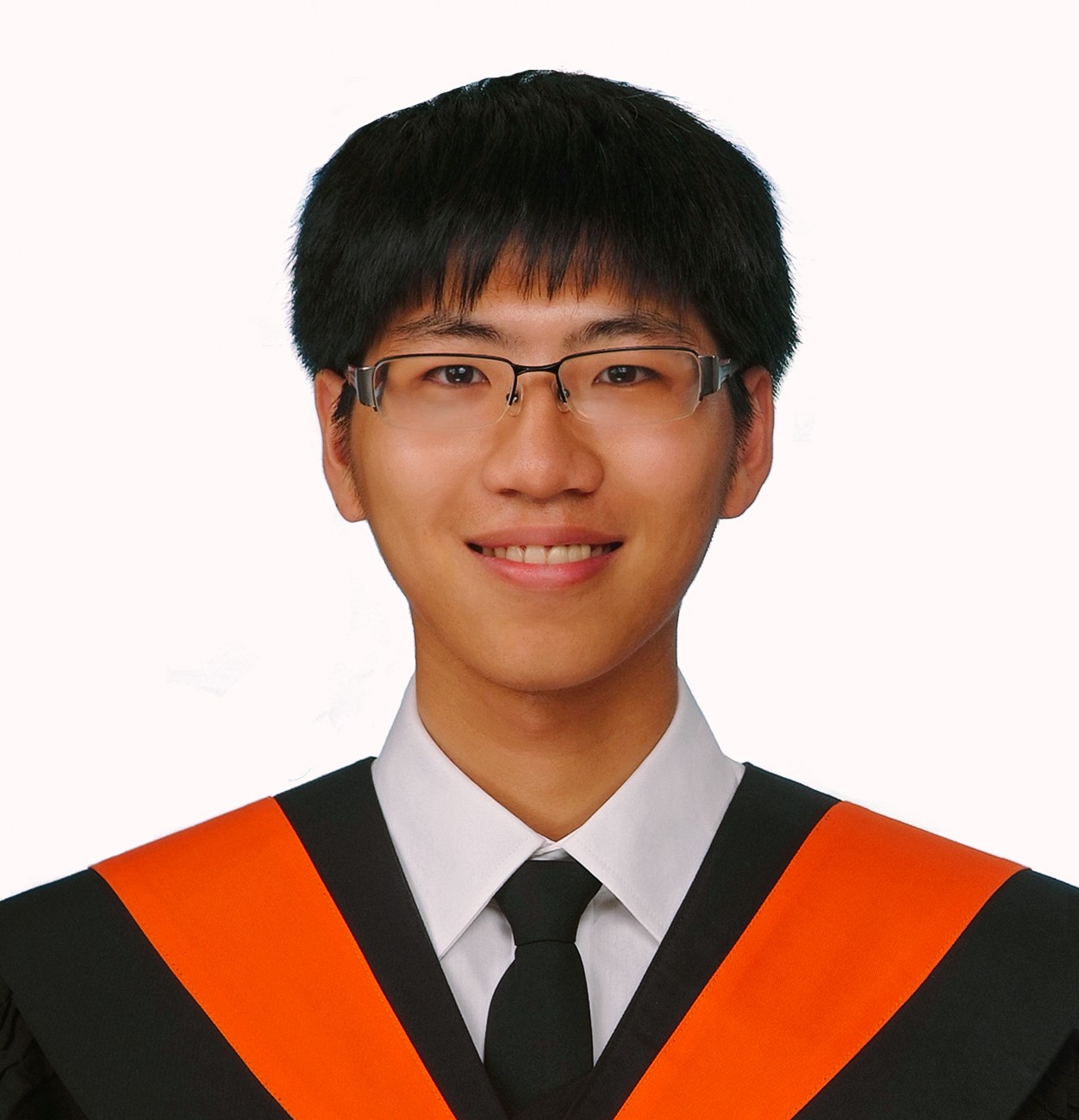}}]{Wei-han Chen} received his B.S. degree in Chemical Engineering from National Taiwan University, Taipei, in 2015, and the M. Eng. degree in Chemical and Biomolecular Engineering from Cornell University, Ithaca, NY, USA, in 2018. He is currently working toward the M. S. degree at Cornell University, Ithaca, NY, USA. His research interests include machine learning and model predictive control.
\end{IEEEbiography}

\vspace{-2cm}

\begin{IEEEbiography}[{\includegraphics[width=1in,clip,keepaspectratio]{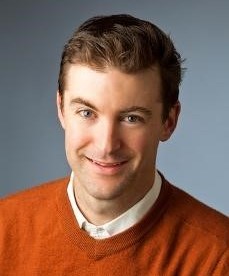}}]{Abraham Duncan Stroock} is the Gordon L. Dibble'50 Professor and William C. Hooey Director of Chemical \& Biomolecular Engineering at Cornell University, Ithaca, NY. After completing a bachelor's degree in Physics at Cornell, he earned a master's degree at the University of Paris VI and XI in Solid State Physics. He then pursue a PhD in the Chemistry department at Harvard University with George Whitesides. In 2003, he joined the faculty of Cornell University. His work involves the development of microfluidic and MEMS systems for the study of biophysics, thermodynamics, and transport phenomena. His lab has focused recently on the development of tools for quantitative measurements in plants. He has been recognized by MIT Technology Review's TR35 list of top innovators and an NSF CAREER Award (2008).
\end{IEEEbiography}

\vspace{-2cm}

\begin{IEEEbiography}[{\includegraphics[width=1in,clip,keepaspectratio]{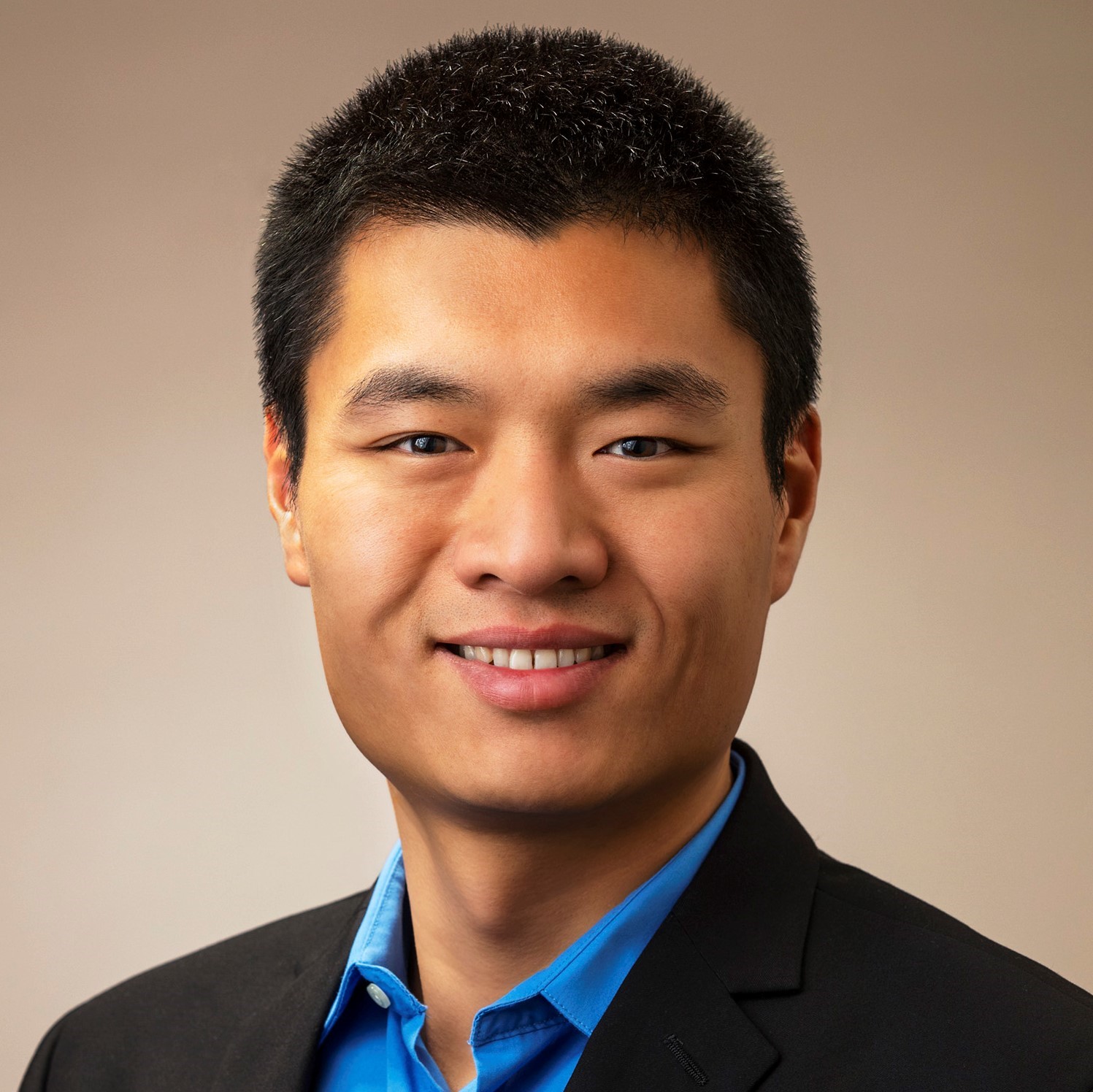}}]{Fengqi You} (M'17) is the Roxanne E. and Michael J. Zak Professor at Cornell University, and is affiliated with the Graduate Fields of Chemical Engineering, Electrical and Computer Engineering, Systems Engineering, Operations Research and Information Engineering, Mechanical Engineering, Civil and Environmental Engineering, and Applied Mathematics. He also serves as Chair of Cornell Systems Engineering PhD Studies and Associate Director of Cornell Energy Systems Institute. He was on the faculty of Northwestern University from 2011 to 2016, and worked at Argonne National Laboratory as an Argonne Scholar from 2009 to 2011. His research focuses on novel computational models, optimization algorithms, statistical machine learning methods, and multi-scale systems analytics tools for process manufacturing, digital agriculture, energy systems, and sustainability. He has published more than 120 peer-reviewed journal articles, and has an h-index of 51. Fengqi earned a B.Eng. from Tsinghua University, and received his Ph.D. from Carnegie Mellon University.
\end{IEEEbiography}

\newpage

\newpage

\noindent \large{\textbf{Supplementary Document for ``Robust Model Predictive Control of Irrigation Systems with Active Uncertainty Learning and Data Analytics"}}

\vspace{0.6cm}

This supplementary document includes fundamentals of support vector clustering (SVC) and proofs of Theorem 1 and Theorem 3.

\section*{Part I: Fundamentals of Support Vector Clustering (SVC)}
SVC maps data $\bold{w}$ into a high-dimensional feature space based on a nonlinear mapping $\boldsymbol{\phi}(\bold{w})$. Given $N$ data samples $\{\bold{w}^{(1)},\cdots,\bold{w}^{(N)}\}$, a circle with minimal volume is sought in the feature space, which encompasses most data samples. This is achieved by solving the following problem:
\begin{equation}
\begin{split}
\min_{\bold{c},R,\boldsymbol{\zeta}}&\ R^2 + {1 \over N\nu}\sum_{i=1}^N \zeta_i\\
\mathrm{s.t.} &\ \| \boldsymbol{\phi}(\bold{w}^{(i)}) - \bold{c} \|^2 \le R^2 + \zeta_i,\ i=1,\cdots,N \\
              &\ \zeta_i \ge 0,\ i=1,\cdots,N
\end{split}
\label{eq:SVC_softmargin}
\end{equation}
where $\bold{c}$ and $R$ are the center and the radius of the circle. Nonnegative auxiliary variables $\{\zeta_i \ge 0\}$ are introduced to indicate data violations of the circle. The objective in (\ref{eq:SVC_softmargin}) aims to simultaneously minimize the volume of the circle and penalize outliers with $\zeta_i > 0$. The regularization parameter $0 < \nu < 1$ is responsible for balancing between two discrepant goals. The dual problem of (\ref{eq:SVC_softmargin}) is essentially a QP \cite{ben01support}:
\begin{equation}
\begin{split}
\max_{\boldsymbol{\alpha}}&\ -\sum_{i=1}^N \sum_{j=1}^N \alpha_i \alpha_j K(\bold{w}^{(i)}, \bold{w}^{(j)}) + \sum_{i=1}^N \alpha_i K(\bold{w}^{(i)}, \bold{w}^{(i)}) \\
\mathrm{s.t.} &\ 0 \le \alpha_i \le 1/N\nu,\ i=1,\cdots,N \\
&\ \sum_{i=1}^N \alpha_i = 1
\end{split}
\label{eq:dualQP}
\end{equation}
where $\{\alpha_i\}$ are Lagrange multipliers, and $K(\bold{w}^{(i)}, \bold{w}^{(j)}) = \boldsymbol{\phi}(\bold{w}^{(i)})^{\rm T}\boldsymbol{\phi}(\bold{w}^{(j)})$ stands for the \textit{kernel function}, which is the inner product between two nonlinear mappings. For data-driven robust optimization, WGIK is proposed by \cite{shang17data}:
\begin{equation}
K(\bold{w},\bold{v}) = \delta - \|\bold{Q}(\bold{w} - \bold{v})\|_1,
\label{eq:WGIK}
\end{equation}
where $\delta$ is a sufficiently large number, $\bold{Q} = \boldsymbol{\Sigma}^{-1/2}$ and $\boldsymbol{\Sigma}$ is the covariance of $\bold{w}$. After solving (\ref{eq:dualQP}), the enclosing circle in the feature space can be established, which yields the data-driven uncertainty set in the data space:
\begin{equation}
\begin{split}
\mathcal{D} & = \left \{ \bold{w} \left | \|\boldsymbol{\phi}(\bold{w}) - \bold{c}\|^2 \le R^2 \right . \right \} \\
& =  \left \{ \bold{w} \left | \sum_{i \in \mathrm{SV}} \alpha_i \|\bold{Q}(\bold{w} - \bold{w}^{(i)})\|_1 \le \theta \right . \right \}
\label{eq:uncertaintyset0}
\end{split}
\end{equation}
where
\begin{gather}
{\rm SV} = \left \{ i \left | \alpha_i > 0,\ 1 \le i \le N \right . \right \}, \\
\theta = \sum_{i \in \mathrm{SV}} \alpha_i \|\bold{Q}(\bold{w}^{(i')} - \bold{w}^{(i)})\|_1,\ i' \in \mathrm{BSV}, \\
{\rm BSV} = \left \{ i \left | 0 < \alpha_i < 1/N\nu,\ 1 \le i \le N \right . \right \}.
\end{gather}

\section*{Part II: Proof of Theorem 1}
It holds that:
\begin{equation}
\begin{split}
&\ \mathbb{P} \left \{ \bold{w} \in \mathcal{W} \right \} \\
= &\ \mathbb{P} \left \{ \boldsymbol{\eta} - \boldsymbol{\xi} \in \mathcal{D}_{\boldsymbol{\eta}} + (-\mathcal{D}_{\boldsymbol{\xi}}(\hat{\bold{p}})) \right \} \\
\ge &\ \mathbb{P} \left \{ \boldsymbol{\eta} \in \mathcal{D}_{\boldsymbol{\eta}}\ {\rm and}\ \boldsymbol{\xi} \in \mathcal{D}_{\boldsymbol{\xi}}(\hat{\bold{p}}) \right \} \\
\ge &\ \mathbb{P} \left \{ \boldsymbol{\eta} \in \mathcal{D}_{\boldsymbol{\eta}} \right \} + \mathbb{P} \left \{ \boldsymbol{\xi} \in \mathcal{D}_{\boldsymbol{\xi}}(\hat{\bold{p}}) \right \} - 1
\end{split}
\end{equation}
where the last inequality is due to a well-known probability inequality. Therefore, the simultaneous occurrence of two events $\mathbb{P} \left \{ \boldsymbol{\eta} \in \mathcal{D}_{\boldsymbol{\eta}}  \right \} \ge 1 - \epsilon_1$ and $\mathbb{P} \left \{ \boldsymbol{\xi} \in \mathcal{D}_{\boldsymbol{\xi}}(\hat{\bold{p}})  \right \} \ge 1 - \epsilon_2$ indicates that the event
\begin{equation}
\mathbb{P} \left \{ \boldsymbol{\eta} - \boldsymbol{\xi} \in \mathcal{D}_{\boldsymbol{\eta}} + (-\mathcal{D}_{\boldsymbol{\xi}}(\hat{\bold{p}})) \right \} \ge 1 - \epsilon_1 - \epsilon_2 = 1 - \epsilon
\end{equation}
occur consequently. Therefore, we have
\begin{equation}
\begin{split}
&\ \mathbb{P}_{\mathcal{W}} \left \{ \mathbb{P} \left \{ \boldsymbol{\eta} - \boldsymbol{\xi} \in \mathcal{D}_{\boldsymbol{\eta}} + (-\mathcal{D}_{\boldsymbol{\xi}}(\hat{\bold{p}})) \right \} \ge 1 - \epsilon \right \} \\
\ge &\ \mathbb{P}_{\mathcal{W}} \left \{ \mathbb{P} \left \{ \boldsymbol{\eta} \in \mathcal{D}_{\boldsymbol{\eta}}  \right \} \ge 1 - \epsilon_1\ \text{and} \ \mathbb{P} \left \{ \boldsymbol{\xi} \in \mathcal{D}_{\boldsymbol{\xi}}(\hat{\bold{p}}) \right \} \ge 1 - \epsilon_2 \right \} \\
\ge &\ \mathbb{P}_{\mathcal{D}_{\boldsymbol{\eta}}} \left \{ \mathbb{P} \left \{ \boldsymbol{\eta} \in \mathcal{D}_{\boldsymbol{\eta}}  \right \} \ge 1 - \epsilon_1 \right \} \\
&\ + \mathbb{P}_{\mathcal{D}_{\boldsymbol{\xi}}(\hat{\bold{p}})} \left \{ \mathbb{P} \left \{ \boldsymbol{\xi} \in \mathcal{D}_{\boldsymbol{\xi}}(\hat{\bold{p}})  \right \} \ge 1 - \epsilon_2 \right \} - 1\\
= &\ 1 - \beta_1 - \beta_2 \\
= &\ 1 - \beta
\end{split}
\end{equation}
This completes the proof.

\section*{Part III: Proof of Theorem 3}
By introducing auxiliary variable $\{ \boldsymbol{\rho}_i,\ i \in {\rm SV} \}$, we first recast the primal optimization problem as:
\begin{equation}
\begin{split}
\max_{\boldsymbol{\xi}^+, \boldsymbol{\xi}^-, \boldsymbol{\rho}_i} &\ \bold{a}^{\rm T} \boldsymbol{\xi}^+ + \bold{b}^{\rm T} \boldsymbol{\xi}^- \\
{\rm s.t.}\ \ & \sum_{i \in {\rm SV}}\alpha_i \cdot \boldsymbol{\rho}_i^{\rm T}\bold{1} \le \theta \\
            & - \boldsymbol{\rho}_i \le \bold{Q}( \boldsymbol{\xi}^+ - \boldsymbol{\xi}^- - \bar{\boldsymbol{\xi}}^{(i)} ) \le \boldsymbol{\rho}_i,\ \forall i \in {\rm SV} \\
            & \bold{0} \le \boldsymbol{\xi}^+,\boldsymbol{\xi}^- \le \bold{1}
\end{split}
\label{eq:worstcase1}
\end{equation}
The associated Lagrangian can be written as:
\begin{equation}
\begin{split}
& L(\boldsymbol{\xi}^+, \boldsymbol{\xi}^-, \boldsymbol{\rho}_i, \bold{r}, \bold{s} , \boldsymbol{\mu}_i, \boldsymbol{\lambda}_i, k) \\
= &\ \bold{a}^{\rm T} \boldsymbol{\xi}^+ + \bold{b}^{\rm T} \boldsymbol{\xi}^- + \bold{r}^{\rm T} (\bold{1} - \boldsymbol{\xi}^+) + \bold{s}^{\rm T} (\bold{1} - \boldsymbol{\xi}^-) \\
  &\ + \sum_{i \in \text{SV}}\boldsymbol{\mu}_i^{\rm T} \left [ \boldsymbol{\rho}_i - \bold{Q}(\boldsymbol{\xi}^+ - \boldsymbol{\xi}^- - \bar{\boldsymbol{\xi}}^{(i)} ) \right ] \\
  &\ + \sum_{i \in \text{SV}}\boldsymbol{\lambda}_i^{\rm T} \left [ \boldsymbol{\rho}_i + \bold{Q}(\boldsymbol{\xi}^+ - \boldsymbol{\xi}^- - \bar{\boldsymbol{\xi}}^{(i)} ) \right ] \\
  &\ + k \left [ \theta - \sum_{i \in {\rm SV}}\alpha_i \cdot \boldsymbol{\rho}_i^{\rm T}\bold{1} \right ]
\end{split}
\end{equation}
where $\bold{r} \ge \bold{0}, \bold{s} \ge \bold{0} , \boldsymbol{\mu}_i \ge \bold{0}, \boldsymbol{\lambda}_i \ge \bold{0}, k\ge 0$ are the Lagrange multipliers. Therefore, the dual function is given by:
\begin{equation}
\begin{split}
&\ g(\bold{r}, \bold{s}, \boldsymbol{\mu}_i, \boldsymbol{\lambda}_i, k) \\
= &\ \max_{\boldsymbol{\xi}^+ \ge \bold{0}, \boldsymbol{\xi}^- \ge \bold{0}, \boldsymbol{\rho}_i} L(\boldsymbol{\xi}^+, \boldsymbol{\xi}^-, \boldsymbol{\rho}_i, \bold{r}, \bold{s} , \boldsymbol{\mu}_i, \boldsymbol{\lambda}_i, k) \\
= &\ \sum_{i \in {\rm SV}} (\boldsymbol{\mu}_i - \boldsymbol{\lambda}_i)^{\rm T} \bold{Q} \bar{\boldsymbol{\xi}}^{(i)} + (\bold{r} + \bold{s})^{\rm T}\bold{1} + k\theta
\end{split}
\end{equation}
when complementary conditions
\begin{gather}
\sum_{i \in {\rm SV}} \bold{Q}(\boldsymbol{\mu}_i - \boldsymbol{\lambda}_i) + \bold{r} \ge \bold{a}, \\
\sum_{i \in {\rm SV}} \bold{Q}(\boldsymbol{\lambda}_i - \boldsymbol{\mu}_i) + \bold{s} \ge \bold{b}, \\
\boldsymbol{\lambda}_i + \boldsymbol{\mu}_i = k \cdot \alpha_i \cdot \bold{1},\ \forall i \in {\rm SV} \end{gather}
are satisfied. Otherwise, $g(\boldsymbol{\pi}_+, \boldsymbol{\pi}_-, , \boldsymbol{\mu}_i, \boldsymbol{\lambda}_i, k) = +\infty$. This yields the dual optimization problem. Due to the compactness of the feasible region of the primal problem as well as the strong duality of linear programs, the objective values of the primal and the dual are identical. This completes the proof.

\ifCLASSOPTIONcaptionsoff
  \newpage
\fi


\begin{thebibliography}{1}
\bibitem{grafton18paradox}
R.~Q. Grafton, J. Williams, C.~J. Perry, F. Molle, C. Ringler, P. Steduto, B. Udall, S.~A. Wheeler, Y. Wang, D. Garrick, and others, ``The paradox of irrigation efficiency," \emph{Science}, vol. 361, no. 6404, pp. 748--750, 2018.

\bibitem{water09united}
U.~N. Water, ``The United Nations world water development report 3?Water in a changing world," United Nations Educational Scientific and Cultural Organization, Paris, 2009.

\bibitem{kim08remote}
Y. Kim, R.~G. Evans, and W.~M. Iversen, ``Remote sensing and control of an irrigation system using a distributed wireless sensor network," \emph{IEEE T. Instrume. Meas.}, vol. 57, no. 7, pp. 1379--1387, 2008.

\bibitem{zhao07automatic}
Y. Zhao, C. Bai, and B. Zhao, ``An automatic control system of precision irrigation for city greenbelt," in \emph{2nd IEEE C. Ind. Elect. Appl.}, May 2007, pp. 2013--2017.

\bibitem{qin03survey}
S.~J. Qin and T.~A. Badgwell, ``A survey of industrial model predictive control technology," \emph{Control Eng. Pract.}, vol. 11, no. 7, pp. 733--764, 2003.

\bibitem{rawlings09model}
J.~B. Rawlings and D.~Q. Mayne, \emph{Model Predictive Control: Theory and Design}, Nob Hill Pub., 2009.

\bibitem{chu2015model}
Y.~Chu and F.~You, ``Model-based integration of control and operations: Overview, challenges, advances, and opportunities, \emph{Comput. Chem. Eng.}, vol. 83, pp. 2--20, 2015.

\bibitem{mccarthy14simulation}
A.~C. McCarthy, N.~H. Hancock, and S.~R. Raine, ``Simulation of irrigation control strategies for cotton using Model Predictive Control within the VARIwise simulation framework," \emph{Comput. Electron. Agr.}, vol. 101, pp. 135--147, 2014.


\bibitem{delgoda16irrigation}
D. Delgoda, H. Malano, S.~K. Saleem, and M.~N. Halgamuge, ``Irrigation control based on model predictive control (MPC): Formulation of theory and validation using weather forecast data and AQUACROP model," \emph{Environ. Modell. Softw.}, vol. 78, pp. 40--53, 2016.

\bibitem{rosolia18data}
U. Rosolia, X. Zhang, and F. Borrelli, ``Data-driven predictive control for autonomous systems," \emph{Annual Review of Control, Robotics, and Autonomous Systems}, vol. 1, pp. 259--286, 2018.

\bibitem{ning17data}
C. Ning and F. You, ``Data-driven adaptive nested robust optimization: General modeling framework and efficient computational algorithm for decision making under uncertainty," \emph{AIChE J.}, vol. 63, no. 9, pp. 3790--3817, 2017.

\bibitem{shang17data}
C. Shang, X. Huang, and F. You, ``Data-driven robust optimization based on kernel learning," \emph{Comput. Chem. Eng.}, vol. 106, pp. 464--479, 2017.

\bibitem{ooi08systems}
S.~K. Ooi, I. Mareels, N. Cooley, G. Dunn, and G. Thoms, ``A systems engineering approach to viticulture on-farm irrigation," in \emph{Proceedings of the 17th IFAC World Congress}, Jul. 2008, pp. 9569-9574.

\bibitem{oldewurtel14stochastic}
F. Oldewurtel, C.~N. Jones, A. Parisio, and M. Morari, ``Stochastic model predictive control for building climate control," \emph{IEEE T. Contr. Syst. T.}, vol. 22, no. 3, pp. 1198--1205, 2014.

\bibitem{zong12application}
Y. Zong, D. Kullmann, A. Thavlov, O. Gehrke, and H.~W. Bindner, ``Application of model predictive control for active load management in a distributed power system with high wind penetration," \emph{IEEE T. Smart Grid}, vol. 3, no. 2, pp. 1055-1062, 2012.

\bibitem{cohen07hydraulic}
S. Cohen, A. Naor, J. Bennink, A. Grava, and M. Tyree, ``Hydraulic resistance components of mature apple trees on rootstocks of different vigours," \emph{J. Exp. Bot.}, vol. 58, no. 15--16, pp. 4213--4224, 2007.

\bibitem{fermandez97a}
R.~T. Fernandez, R.~L. Perry, and J.~A. Flore, ``Drought response of young apple trees on rhree rootstocks: Growth and development," \emph{J. Am. Soc. Hortic. Sci.}, vol. 122, no. 1, pp. 14--19, 1997.

\bibitem{fermandez97b}
R.~T. Fernandez, R.~L. Perry, and J.~A. Flore, ``Drought response of young apple trees on rhree rootstocks. II. Gas exchange, chlorophyll fluorescence, water relations, and leaf abscisic acid," \emph{J. Am. Soc. Hortic. Sci.}, vol. 122, no. 6, pp. 841--848, 1997.

\bibitem{mao18soil}
Y. Mao, S. Liu, J. Nahar, J. Liu, and F. Ding, ``Soil moisture regulation of agro-hydrological systems using zone model predictive control," \emph{Comput. Electron. Agr.}, vol. 154, pp. 239--247, 2018.

\bibitem{saltik18outlook}
M.~B. Salt{\i}k, L. {\"O}zkan, J.~H. Ludlage, S. Weiland, and P.~M. Van den Hof, ``An outlook on
robust model predictive control algorithms: Reflections on performance and computational aspects," \emph{J. Process Contr.}, vol. 61, pp. 77--102, 2018.

\bibitem{chen14post}
J.~Chen, F.~P. Brissette, and Z.~Li, ``Postprocessing of ensemble weather forecasts using a stochastic weather generator," \emph{Mon. Weather Rev.}, vol. 142, pp. 1106--1124, 2014.

\bibitem{kirtman14north}
B.~P. Kirtman, D. Min, J.~M. Infanti et al., ``The north American multimodel ensemble: Phase-1 seasonal-to-interannual prediction; Phase-2 toward developing intraseasonal prediction," \emph{B. Am. Meterol. Soc.}, vol. 95, pp. 585--601, 2014.

\bibitem{schepen18bayesian}
A. Schepen, T. Zhao, Q.~J. Wang, and D.~E. Robertson, ``A Bayesian modelling method for post-processing daily sub-seasonal to seasonal rainfall forecasts from global climate models and evaluation for 12 Australian catchments," \emph{Hydrol. Earth. Syst. Sc.}, vol. 22, no. 2, 1615--1628, 2018.

\bibitem{Schwanenberg15short}
D. Schwanenberg, F. M. Fan, S. Naumann, J. I. Kuwajima, R. A. Montero, and A. A. Dos Reis, ``Short-term reservoir optimization for flood mitigation under meteorological and hydrological forecast uncertainty," \emph{Water Resour. Manag.}, vol. 29, no. 5, pp. 1635--1651, 2015.

\bibitem{zhao13unified}
C. Zhao and Y. Guan, ``Unified stochastic and robust unit commitment," \emph{IEEE T. Power Syst.}, vol. 28, no. 3, pp. 3353--3361, 2013.

\bibitem{ning19data}
C. Ning and F. You, ``Data-driven adaptive robust unit commitment under wind power uncertainty: A Bayesian nonparametric approach," \emph{IEEE T. Power Syst.}, vol. 34, pp. 2409--2418, 2019.

\bibitem{vij10assessing}
M.~M. Vijayalakshmi, V.~N. Mariappan, ``Assessing a suitable method in estimating evapotranspiration for crop water requirement," in \emph{Recent Advances in Space Technology Services and Climate Change}, Nov. 2010, pp. 240--243.

\bibitem{tian12forecasting}
D. Tian and C.~J. Martinez, ``Forecasting reference evapotranspiration using retrospective forecast analogs in the southeastern united states," \emph{J. Hydrometeorol}, vol. 13, pp. 1874--1892, 2012.

\bibitem{iowa}
``Meteogram Generator," Dept. of Geological and Atmospheric Sciences, Iowa State Univ. Available: \url{http://www.meteor.iastate.edu/~ckarsten/bufkit/image_loader.phtml}.

\bibitem{chen07robust}
X. Chen, M. Sim, and P. Sun, ``A robust optimization perspective on stochastic programming," \emph{Oper. Res.}, vol. 55, no. 6, pp. 1058--1071, 2007.

\bibitem{ning18data}
C. Ning and F. You, ``Data-driven decision making under uncertainty integrating robust optimization with principal component analysis and kernel smoothing methods," \emph{Comput. Chem. Eng.}, vol. 112, pp. 190--210, 2018.

\bibitem{hong16approximating}
L.~J. Hong, Z. Huang, and H. Lam, ``Approximating data-driven joint chance-constrained programs via uncertainty set construction," in \emph{Wint. Simul. C. Proc.}, pp. 389--400, 2016.

\bibitem{shang18data}
C. Shang and F. You, ``A data-driven robust optimization approach to scenario-based stochastic model predictive control," \emph{J. Process Contr.}, vol. 75, pp. 24--39, 2019.

\bibitem{goulart06optimization}
P.~J. Goulart, E.~C. Kerrigan, and J.~M. Maciejowski, ``Optimization over state feedback policies for robust control with constraints," \emph{Automatica}, vol. 42, no. 4, pp. 523--533, 2006.

\bibitem{bental04adjustable}
A. Ben-Tal, A. Goryashko, E. Guslitzer, and A. Nemirovski, ``Adjustable robust solutions of uncertain linear programs," \emph{Math. Program.}, vol. 99, no. 2, pp. 351--376, 2004.

\bibitem{ning17data1}
C. Ning and F. You, ``A data-driven multistage adaptive robust optimization framework for planning and scheduling under uncertainty," \emph{AIChE J.}, vol. 63, no. 10, pp. 4343--4369.

\bibitem{georghiou15generalized}
A. Georghiou, W. Wiesemann, and D. Kuhn, ``Generalized decision rule approximations for stochastic programming via liftings," \emph{Math. Program.}, vol. 152, no. 1-2, pp. 301--338, 2015.

\bibitem{zhang17robust}
X. Zhang, M. Kamgarpour, A. Georghiou, P. Goulart, and J. Lygeros, ``Robust optimal control with adjustable uncertainty sets," \emph{Automatica}, vol. 75, pp. 249--259, 2017.

\bibitem{hargreaves85reference}
G.~H. Hargreaves and Z.~A. Samani, ``Reference crop evapotranspiration from temperature," \emph{Appl. Eng. Agric.}, vol. 1, no. 2, pp. 96--99, 1985.




\bibitem{grant08matlab}
M. Grant, S. Boyd, and Y. Ye, ``CVX: MATLAB software for disciplined convex programming," 2008.

\bibitem{jensen80design}
M.~E. Jensen, ``Design and operation of farm irrigation systems," American Society of Agricultural Engineers, Technical Report, 1980.

\bibitem{goodwin95irrigation}
I. Goodwin, \textit{Irrigation of Vineyards: A Winegrape Grower's Guide to Irrigation Scheduling and Regulated Deficit Irrigation}, Institute of Sustainable Irrigated Agriculture, 1995.


\end{thebibliography}

\begin{thebibliography}{1}

\bibitem{ben01support}
A. Ben-Hur, D. Horn, H.~T. Siegelmann, and V. Vapnik, ``Support vector clustering," \emph{Journal of Machine Learning Research}, vol. 2, no. Dec, pp. 125--137, 2001.

\bibitem{shang17data}
C. Shang, X. Huang, and F. You, ``Data-driven robust optimization based on kernel learning," \emph{Computers \& Chemical Engineering}, vol. 106, pp. 464--479, 2017.

\end{thebibliography}
\end{document}